\documentclass[journal]{new-aiaa}
\usepackage[utf8]{inputenc}

\usepackage{graphicx}
\usepackage{amsmath}
\usepackage{subcaption}
\usepackage[version=4]{mhchem}
\usepackage{siunitx}
\usepackage{longtable,tabularx}
\usepackage[table]{xcolor}
\usepackage[table]{xcolor}
\usepackage{bm}
\usepackage{multirow}
\usepackage{graphicx}
\usepackage{makecell}

\title{Sweep Angle Effects of Flow Over a Seal Whisker-Inspired Undulated Cylinder}
\author{Trevor K. Dunt\footnote{PhD Candidate, Nuclear Engineering \& Engineering Physics. Corresponding author: \texttt{dunt@wisc.edu}}}
\affil{University of Wisconsin--Madison, Madison, WI, 53706}
\author{Christin T. Murphy\footnote{Head, Bioinspired Research and Development Lab}}
\affil{Naval Undersea Warfare Center--Newport, Newport, RI, 02841}
\author{Ond\v rej Fer\v c\'ak\footnote{PhD Candidate, Mechanical \& Materials Engineering}}
\author{Ra\'ul Bayo\'an Cal\footnote{Professor, Mechanical \& Materials Engineering}}
\affil{Portland State University, Portland, OR, 97201}
\author{Jennifer A. Franck\footnote{Associate Professor, Mechanical Engineering}}
\affil{University of Wisconsin--Madison, Madison, WI, 53706}

\begin{document}

\maketitle
\begin{abstract}
Flow over a seal whisker-inspired undulated cylinder at swept back angles is computationally investigated, comparing the vortex shedding, forces, and wake characteristics to those of an equivalent smooth geometry. Numerous prior studies have demonstrated that undulated cylinders can reduce mean drag and unsteady lift oscillations; however, none have isolated the effects of the sweep angle resulting from whisker positioning in flow. Inspired by the active control seals exert over their whiskers while navigating and sensing in unsteady aquatic environments, this study investigates how such orientation influences the hydrodynamic performance of the geometry. 

Simulations are performed of flow across a rigid, infinite-span, undulated cylinder at sweep angles from 0 to 60$\bm{^\circ}$ and at Reynolds numbers of 250 and 500. At zero sweep, the undulated cylinder breaks up coherent two-dimensional vortices, having the effect of reducing drag by 11.4\% and root mean square lift by 90.8\% compared to a smooth elliptical cylinder. With sweep added, the prominence of spanwise vortex breakup and force suppression is reduced, approximating flow over smooth ellipse geometry as sweep increases. At low sweep angles of 15 and 30 degrees, lift is still suppressed by 72.4\% and 47.6\% while drag results in a smaller difference of 5.7 and 1.6\% reduction from a smooth ellipse. These results reinforce that sweep angle is a significant parameter both mechanically and biologically in the flow physics of whisker-inspired undulated geometries.
\end{abstract}

\section{Introduction}
\label{sec:introduction}
\begin{figure}[hbt!]
\begin{subfigure}{0.49\textwidth}
\centering
\includegraphics[width=0.8\textwidth]{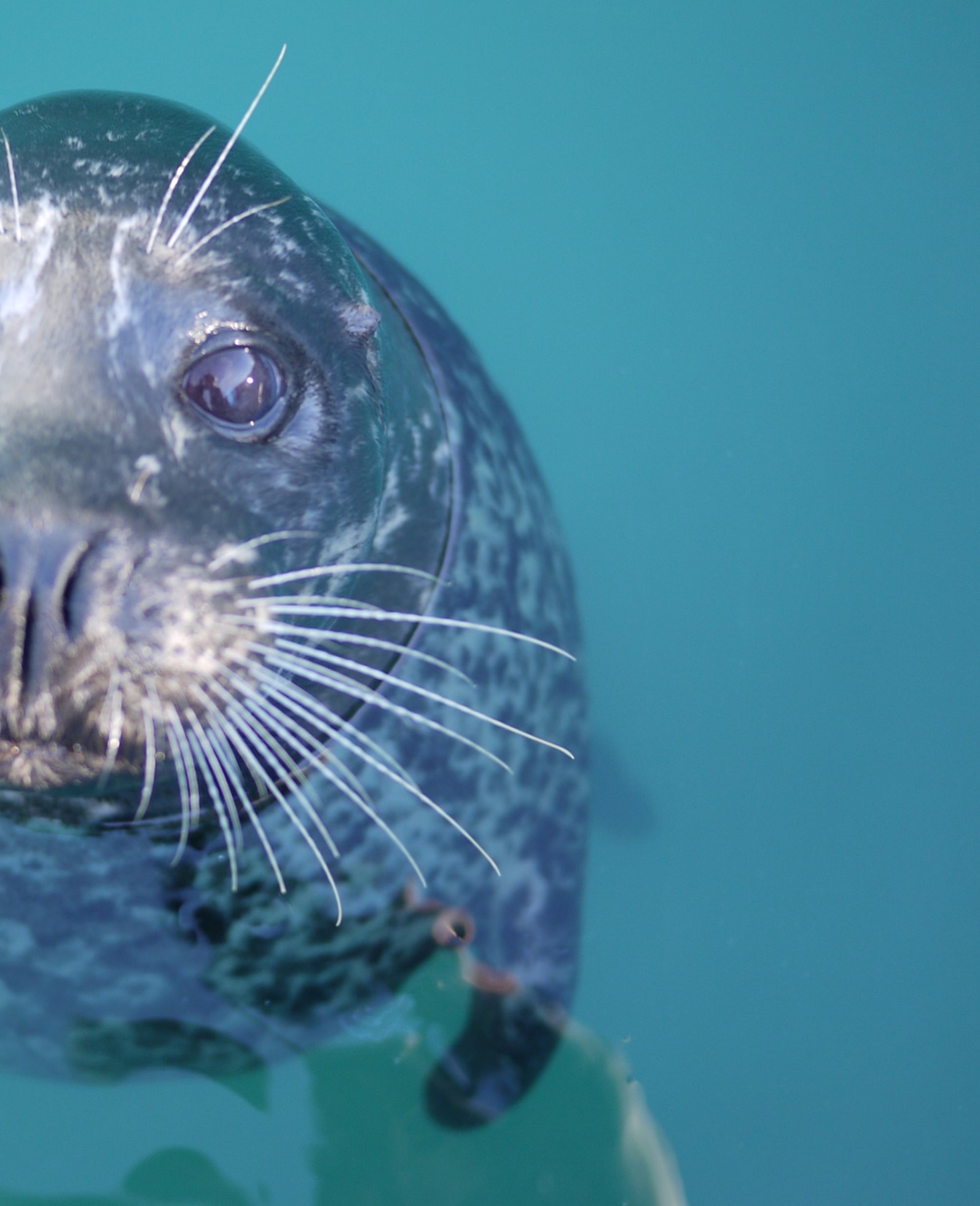}
\caption{Whiskers in swept-back, relaxed position}
\label{fig:flat}
\end{subfigure}
\begin{subfigure}{0.49\textwidth}
\centering
\includegraphics[width=0.8\textwidth]{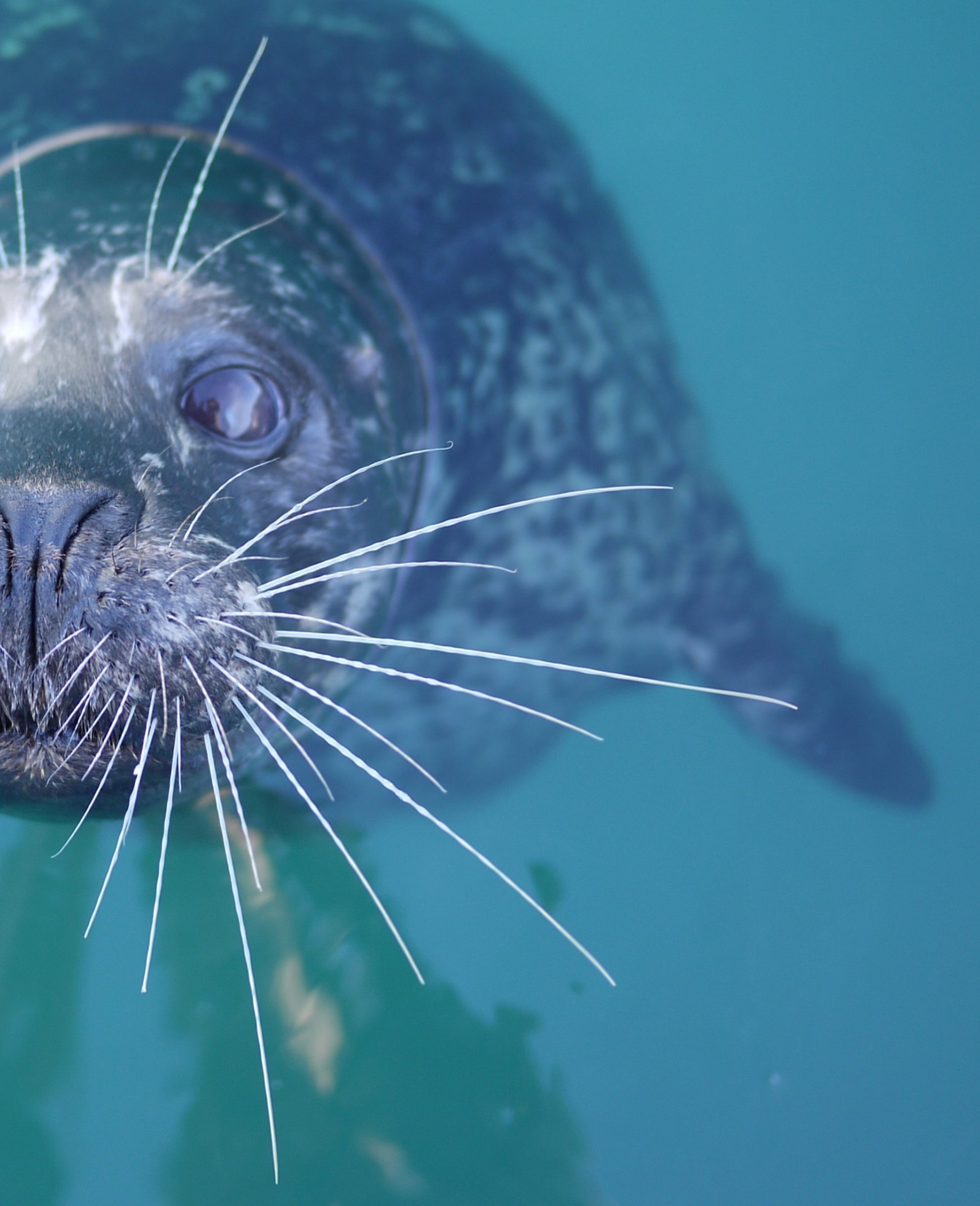}
\caption{Whiskers protracted away from face}
\label{fig:protract}
\end{subfigure}
\caption{Harbor seal Sprouts, UCSC Pinniped Lab demonstrating a range of whisker movement. (National Marine Fisheries Service (NMFS) permit 14535) Panel (b) adapted from \citet{murphy_navy_seals_2021}.}
\label{fig:Seal_photo}
\end{figure}

\begin{figure}[hbt!]
\centering
\begin{minipage}{0.72\textwidth}
\centering
\begin{subfigure}[t]{1\textwidth}
\includegraphics[width=\linewidth]{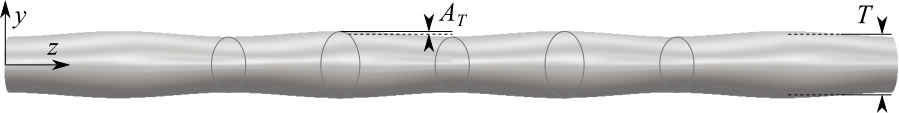}
\caption{Frontal view of whisker thickness profile.}
\label{fig:slice_diagram_a}
\end{subfigure}
\begin{subfigure}[t]{1\textwidth}
\includegraphics[width=\linewidth]{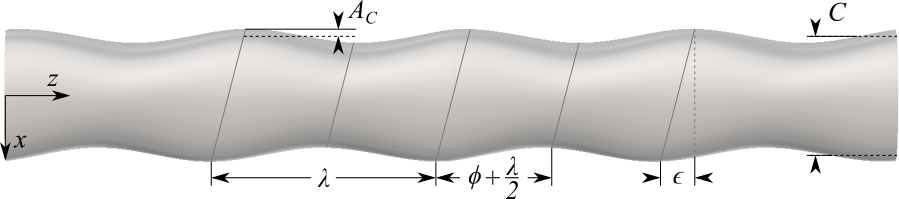}
\caption{Top-down view of whisker chord profile.}
\label{fig:slice_diagram_b}
\end{subfigure}
\end{minipage}
\hfill
\begin{minipage}{0.27\textwidth}
\centering
\vfill
\begin{subfigure}[c]{1\textwidth}
\includegraphics[width=\linewidth]{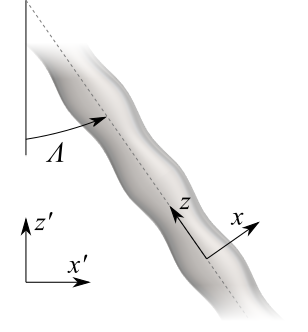}
\caption{Sweep angle definition of rotation about the $y$-axis.}
\end{subfigure}
\vfill
\end{minipage}
\caption{Geometric variables comprising the undulated cylinder model as defined by \citet{lyons_flow_2020}.}
\label{fig:whisker_geom}
\end{figure}

Distinct from the smooth whiskers of other mammals, most species of true seal (such as the harbor seal pictured in Figure~\ref{fig:Seal_photo}) have undulated whiskers \cite{morgenthal_characterization_2025}.
The whisker shape, comprised of two opposing sets of undulations as depicted in Figure~\ref{fig:whisker_geom}, enhances seals' excellent navigation and prey tracking abilities enabled through hydrodynamic sensing \cite{dehnhardt_seal_1998,dehnhardt2001,Murphy2017}. The undulated geometry reduces drag and oscillating lift forces compared to smooth whiskers, improving the effective signal to noise ratio when sensing their environment \cite{hanke_harbor_2010,lyons_flow_2020}. 

Prior research on seal whiskers began primarily with biological and behavioral studies which established that undulated whiskers play a central role in hydrodynamic sensing, enabling seals to detect water motions and track wakes generated by prey \cite{dehnhardt_seal_1998,dehnhardt2001,Murphy2017}. These findings motivated extensive investigation into the physical mechanisms underlying this geometry’s modification of flow compared to canonical bluff bodies \cite{zheng_review_2021}. 
To address these questions, experimental and numerical studies demonstrated that the undulated whisker geometry operates by suppressing vortex-induced vibrations and reduces mean drag and unsteady lift force compared to smooth cylinders by disrupting spanwise vortex coherence \cite{hanke_harbor_2010,witte_wake_2012,wang_wake_2016,liu_phase-difference_2019}. 
Parametric modeling studies were then undertaken to isolate the geometric features most responsible for these effects, identifying wavelength, aspect ratio, curvature, and undulation amplitude as dominant contributors to force reduction and wake modification \cite{lyons_flow_2020,yuasa_simulations_2021,lyons_dye_2021,kamat_wavelength_diameter_2024,dunt_frequency_2024,yuanji_curvature_2025}. 

In parallel with these hydrodynamic investigations, a growing body of work has examined how altered wake structure and force response contribute to the whisker’s sensing function as a biological structure or man-made bioinspired sensor. Experimental and computational studies have demonstrated the ability of whisker-inspired structures to detect and track upstream wakes, often incorporating fluid–structure interaction or imposed upstream disturbances to model realistic sensing scenarios, where whisker motion or environmental disturbances are central to the problem formulation \cite{beem_calibration_2013, beem_wake-induced_2015,kamat_wavelength_diameter_2024,zheng_printed_whisker_sensor_2022,eberhardt_whisker_sensor_2016,geng_whisker_sensor_2025,liu_phase-difference_2019,zhu_FIV_tandem_whiskers_2026,liu_flow-signal-correlation_2022}.
Beyond biological analogs, related flow phenomena have also been observed for non-biological wavy or sinusoidal cylinders, where spanwise geometric modulation disrupts vortex shedding and alters force response \cite{ahmed_transverse_1992,lam_effects_2009,lam_three-dimensional_2004,zhang_piv_2005}. However, such idealized geometries differ substantially from seal whiskers in cross-sectional shape, undulation pattern, and biological scaling.

Despite this extensive background, seal whisker research has almost exclusively considered span-perpendicular inflow. While angle of attack effects under perpendicular flow have been examined \cite{murphy_effect_2013,kim_effect_2017}, the influence of sweep angle, where the flow is no longer perpendicular to the whisker axis, has not been systematically investigated. This gap is particularly relevant given the active and variable orientation of whiskers during swimming and foraging.

As shown by Figure \ref{fig:Seal_photo}, seals exert active musculature control to protract their whiskers \cite{milne_orient_control_2020,adachi_prey_sensors_2022}. In particular, during active foraging, elephant seals have been recorded alternating between protracting their whiskers outward into flow, particularly when prey is present, and sweeping them back against their face \cite{adachi_prey_sensors_2022}. Additionally, individual whiskers exhibit intrinsic curvature \cite{luo_intrinsic_curvature_2023,rinehart_characterization_2017} and are distributed across the face with varying spatial orientation \cite{dougill_whisker_morphology_2023,graff_spatial_arrangement_2024}. 
Combined with head and neck motion during swimming, these factors introduce natural variation in flow direction relative to each whisker, making sweep angle a relevant variable in this system.  Despite this biological relevance, the effects of sweep on seal whiskers have not been previously studied in detail, as it is difficult to capture nuanced whisker data from live seals or excised whiskers \cite{Murphy2017}. Computational simulations, however, provide a valuable tool for isolating specific geometric and orientation variables in seal whisker hydrodynamics that are difficult to capture in experimental measurements or biological samples. In addition, computational studies of whisker-inspired geometries can inform engineering applications, such as flow sensing, drag reduction, and wake control, where analogous three-dimensional undulations or orientation effects may be exploited.

A three-dimensional surface model of a seal whisker was introduced by \citet{hanke_harbor_2010}, and defined by two angled ellipse cross sections and a distance between them. \citet{lyons_flow_2020} redefined the model into independent and hydrodynamically relevant variables and established the sensitivity of each parameter, finding  wavelength, aspect ratio, and undulation amplitudes to be most impactful on forces. Follow-up studies performed analysis focusing on specific geometric parameters\cite{yuasa_simulations_2021,lyons_dye_2021,lyons_wavelength_2022, dunt_frequency_2024,kamat_wavelength_diameter_2024,yuanji_curvature_2025}, documenting the capabilities of the whisker-inspired geometry such that they may be extended to bio-inspired engineering applications. 
Angle of attack effects (rotation about the $\mathit{z}$-axis) have been investigated previously for span-perpendicular flow, finding that reduced forces are most prevalent at zero angle of attack while the flow response increasingly approximates flow over a bluff body as the angle of attack reaches 90 degrees \cite{murphy_effect_2013,wang_flow_2017,kim_effect_2017,geng_angle-of-attack_VIV_2024}. 

Here, sweep angle is defined by $\mathit{\Lambda}$ in Figure~\ref{fig:whisker_geom}, where $\mathit{\Lambda}=0$ corresponds to pure perpendicular flow. Commonly called sweep angle in aerodynamics, alternate terminology such as yaw angle or crossflow are often utilized in bluff body hydrodynamics. Research of crossflow over a smooth circular cylinder include experiments \cite{king_vortex_1977, ramberg_effects_1983, williamson_oblique_1989, snarski_flow_2004, thakur_wake_2004} and simulations \cite{wang_effect_2019, lucor_effects_2003, marshall_wake_2003, thakur_wake_2004, zhao_direct_2009} over a range of Reynolds number regimes. These studies serve as a reference for the expected behavior of bluff bodies at sweep angles, and represent a useful baseline for understanding the comparatively unique flow over the undulated cylinder. These works often invoke the classical independence principle (IP), which estimates forces on a swept body by scaling perpendicular flow values with the normal velocity component. While IP neglects three-dimensional flow effects, it can reasonably reproduce projected-area scaling and associated mean drag trends for some bluff-body geometries at moderate sweep angles. However, IP does not account for changes in wake instability, vortex shedding coherence, or spanwise phase variation, and therefore becomes less reliable for predicting forces by vortex shedding dynamics \cite{ramberg_effects_1983}.

This paper presents simulations of flow over an infinite-span model with sweep angles from 0 to 60 degrees, investigating how sweep impacts the force-reduction properties of an undulated cylinder. The sweep angles considered here are not intended to represent specific biological postures, but rather span a representative range of effective whisker orientations arising from curvature, placement, and motion. The intent is to isolate how sweep modifies force suppression and wake structure, rather than to prescribe a particular behavioral configuration. Section~\ref{sec:methods} introduces the geometry, the numerical model, and the flow parameters, whereas Section~\ref{sec:results} presents results and discussion, and concluding remarks are summarized in Section~\ref{sec:conclusion}. 

\section{Methods}
\label{sec:methods}
This research computes the time-dependent and three-dimensional flow around an undulated cylinder at various inflow sweep angles. The governing equations are the non-dimensional and incompressible Navier-Stokes equations, 

\begin{equation}
\label{NS1:equation}
\frac{\partial\textbf{u}}{\partial t} + \textbf{u} \cdot \nabla\textbf{u} = -\nabla p + \frac{1}{Re} \nabla^{2} \textbf{u}
\end{equation}

\noindent and

\begin{equation}
\label{NS2:equation}
\nabla \cdot \textbf{u} = 0\text{,}
\end{equation}

\noindent where $\textbf{u}$ is the normalized velocity vector, and $p$ and $t$ are non-dimensional pressure and time. Quantities are normalized by the characteristic freestream velocity $U_{\infty}$ and the mean thickness of the cylinder, $T$. The Reynolds number is thus defined as 

\begin{equation}
\label{Re:equation}
Re = \frac{U_{\infty}T}{\nu},
\end{equation}

\noindent where $\nu$ is the kinematic viscosity of the fluid. 

The numerical solution is discretized using \textit{OpenFOAM} \cite{openfoam} second-order finite volume libraries, and solved with a pressure-implicit split-operator (PISO) algorithm. The numerical method follows that of prior work  \cite{lyons_wavelength_2022, yuasa_simulations_2021, dunt_frequency_2024}.

\begin{figure}[hbt!]
\centering
\begin{minipage}{0.56\textwidth}
\centering
\begin{subfigure}[t]{1\textwidth}
\includegraphics[width=\linewidth]{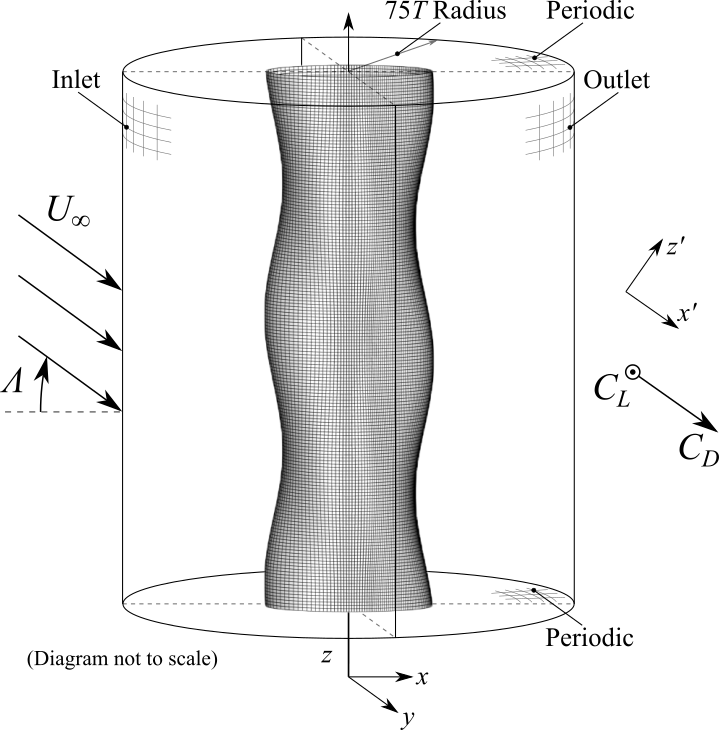}
\caption{The cylindrical computational domain and coordinate systems}
\label{fig:domain}
\end{subfigure}
\end{minipage}
\hspace{5mm}
\begin{minipage}{0.35\textwidth}
\centering
\begin{subfigure}[c]{1\textwidth}
\includegraphics[width=\linewidth]{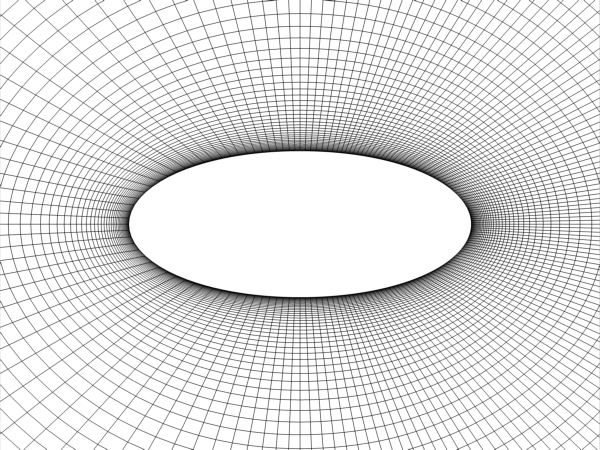}
\caption{Mesh boundary layer region}
\label{fig:mesh_zoom}
\end{subfigure}
\begin{subfigure}[c]{1\textwidth}
\includegraphics[width=\linewidth]{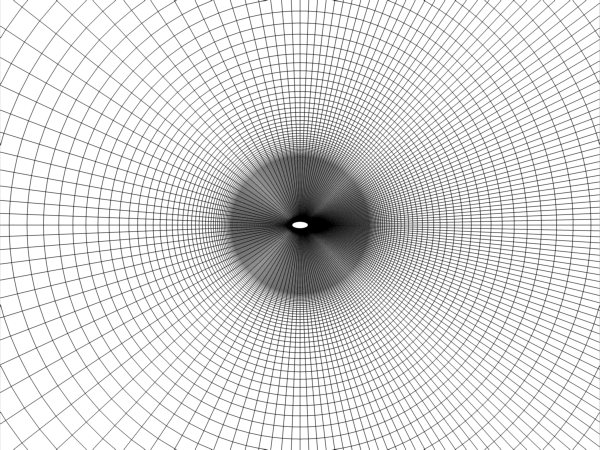}
\caption{Mesh outer region}
\label{fig:mesh_large}
\end{subfigure}
\end{minipage}
\caption{Undulated cylinder computational domain and radial mesh cell distribution.}
\end{figure}

The undulated geometry is comprised of two sets of undulations, one in the chord ($x$-direction), and another in the thickness ($y$-direction), as seen in Figure~\ref{fig:whisker_geom}. The surface undulations are prescribed from the hydrodynamically relevant non-dimensional parameters developed by \citet{lyons_flow_2020} with mean thickness $T = 1$, aspect ratio $\gamma = C/T = 1.919$, wavelength $\lambda/T = 3.434$, transverse amplitude $A_T/T = 0.09$, streamwise amplitude $A_C/T = 0.23$, undulation offset $\epsilon = 0.34$, and undulation symmetry $\phi = 0.015$. These values yield a surface with the same measurements as that originally defined by Hanke et al. which enables comparison to prior results of the same geometry at a similar Reynolds number. The conversion between the two geometric definitions is provided in \citet{lyons_flow_2020}. 

The analytical surface of the whisker is generated and defined using the mesh morphing algorithm described by \citet{yuasa_simulations_2021}. Starting from a circular cylinder computational domain with a radially structured mesh, the algorithm radially translates the surface mesh nodes to morph the cylinder into a whisker geometry with the specified geometric features defined above. The result is a body-fitted, structured mesh of an undulated cylinder that is repeatable and independent of changes to the starting cylinder mesh resolution.

The computational domain, shown in Figure~\ref{fig:domain}, consists of an outer cylinder with a radius of 75$T$, with a body-fixed computational coordinate system defined such that $x$ and $y$ are perpendicular and $x$ is aligned to the major (chord) axis, $y$ with the shorter thickness axis, and $z$ is in the spanwise direction. The outer boundary is divided into an upstream inlet with uniform velocity and zero-gradient pressure and a downstream, zero-gradient velocity and fixed pressure outlet. The model span consists of two wavelengths of the undulated cylinder identified in prior literature as sufficient to resolve the periodic flow structures created \cite{kim_effect_2017}, with periodic boundary conditions in the $z$-direction. Because the present spanwise domain length was selected based on resolution requirements for perpendicular inflow, and no span-length sensitivity study was conducted for strongly swept configurations, the possibility of longer-wavelength instabilities unique to high sweep cannot be fully excluded. However, the observed wake structures and force trends do not exhibit signatures of unresolved large-scale spanwise modes.

Sweep angles are varied by prescribing an angled inlet velocity relative to the body-fixed coordinate system. For the purpose of force decomposition, a rotated, freestream-aligned coordinate system ($x^\prime$, $y^\prime$, $z^\prime$) is introduced where $x^\prime$ is in the freestream direction, $y^\prime$=$y$, and $z^\prime$ is perpendicular to the ($x^\prime$, $y^\prime$) plane. Due to the phase difference in leading and trailing edge undulations, the geometry possesses a directional bias relative to the sign of sweep angle. The positive $\mathit{\Lambda}$ direction is chosen such that angled flow is in line with the angled ellipses that define the whisker surface as visible in Figure~\ref{fig:whisker_geom}. In the current investigation, only the positive (aligned with the undulation offset $\epsilon$) range of sweep angle $\mathit{\Lambda}$ is explored. 

The cylindrical structured mesh shown in Figure~\ref{fig:domain} consists of 170 cells in the spanwise direction resulting in a spanwise resolution of $\Delta z/T = 0.041$, 160 in the azimuthal direction, and 154 in the radial direction which are initially clustered to produce a minimum cell thickness of $\Delta r/T = 0.003$ and radial resolution of $\Delta \theta = 4.01^\circ$. While the mesh and computational domain have been previously validated at zero sweep angle \cite{yuasa_simulations_2021,lyons_wavelength_2022}, a mesh independence study at sweep angle $\mathit{\Lambda} = 60$ is documented in Section~\ref{appA}.

Simulations are performed at Reynolds numbers of 250 and 500, which are biologically relevant based on seal swimming speeds and average whisker diameter \cite{Lesage_classification_1999}. Both values lie within the same low-to-moderate Reynolds number regime commonly studied in whisker-inspired bluff-body flows and have been previously observed to demonstrate two distinct shedding modes from the same geometry \cite{lyons_flow_2020,lyons_dye_2021,yoon_effect_2020,lam_effects_2009,hanke_harbor_2010,lyons_wavelength_2022,yuasa_simulations_2021,dunt_frequency_2024,kim_effect_2017,morrison_simulating_2016}. By examining two Reynolds numbers, the robustness of sweep-induced changes in force suppression and wake structures is explored over these two regimes. Prior simulations show that $Re = 250$ produces clean and repeatable hairpin vortices, whereas at $Re = 500$, these structures begin to break up and transition in the near wake \cite{lyons_flow_2020,lyons_wavelength_2022,yuasa_simulations_2021,dunt_frequency_2024}.Simulation time is specified in non-dimensional convective time units (CTU), or the time it takes for the flow to move one mean thickness at freestream velocity. All simulations begin with an initial transient time of 100 CTU to allow ample time for the flow to fully develop. After this point, a subsequent 400 CTU are simulated and utilized for all averaged quantities.

Forces due to pressure and skin friction are computed along the whisker surface as a function of time, and are used to compute the lift and drag coefficients. As sweep angle rotation occurs in the $xz$ plane, the magnitude of lift forces oscillating parallel to the $y$-axis require no rotation, whereas drag force will shift directions with changes in $\mathit{\Lambda}$ and is thus defined parallel to the freestream flow along $x^\prime$. The drag and lift coefficients are given by

\begin{equation}
\label{drag:equation}
C_D = \frac{2F_D}{\rho U_\infty^2 T L_z} \quad \text{and} \quad C_L = \frac{2F_L}{\rho U_\infty^2 C L_z}
\end{equation}

\noindent where drag force is normalized by the whisker mean thickness $T$ (frontal area in the $yz$ plane), and lift is normalized by the mean chord length $C$ (an average area in $xz$). As freestream velocity magnitude is held constant, increasing sweep angle reduces the velocity component perpendicular to the span which is often responsible for driving dominant forces. As a result, a systematic reduction in force magnitudes occurs when normalizing coefficients with $U_\infty$. To aid in examination of force trends across sweep angle, an alternative normalization based on the perpendicular velocity component $U_{\perp} = U_\infty cos(\mathit{\Lambda})$ is

\begin{equation}
\label{dragnorm:equation}
C_{D\perp} = \frac{2F_D}{\rho U_\perp^2 T L_z} \quad \text{and} \quad C_{L\perp} = \frac{2F_L}{\rho U_\perp^2 C L_z}.
\end{equation}

\noindent As the whisker geometry lacks camber and is symmetric about the $xz$ plane, mean lift is zero and thus the magnitude of oscillations is captured by the root mean square (RMS) of the coefficient over $n$ samples in time with a sampling rate of 50 samples per convective time unit over the 400 CTU statistically developed period,

\begin{equation}
\label{RMS:equation}
C_{L, RMS} = \sqrt{\frac{1}{n} \sum_{i=1} C_{L_i}^2}.
\end{equation}

To provide a baseline comparison, simulations of flow over a circular cylinder and elliptical cylinder of equivalent mean thickness are performed. The elliptical cylinder maintains the same mean aspect ratio as the whisker model removing effects of the streamlined cross-section and serves as a useful intermediate comparison between a circular cylinder and the undulated cylinder.

Shedding frequency is computed from a fast Fourier transform (FFT) of the spanwise-averaged lift coefficient to quantify the dominant unsteady behavior across sweep angles. A reduced frequency $f^*$ is computed as 

\begin{equation}
\label{freq:equation}
f^{*} = \frac{fT}{U_{\infty}}.
\end{equation}

To aid in flow visualization $Q$-criterion is calculated to highlight vortices in the wake by relating the magnitude of local strain rate $\textbf{\textit{S}} = \frac{1}{2}(\nabla\textbf{\textit{u}}+\nabla\textbf{\textit{u}}^T)$ and rotation rate $\bm{\varOmega} = \frac{1}{2}(\nabla\textbf{\textit{u}}-\nabla\textbf{\textit{u}}^T)$ tensors,
\begin{equation}
\label{Q:equation}
Q = \frac{1}{2}(||\bm{\varOmega}||^2-||\textbf{\textit{S}}||^2).
\end{equation}

Lastly, the distribution of energy in the wake is compared via contours of turbulent kinetic energy (TKE), defined by the half-sum of variances $\overline{(\textbf{\textit{u}}^\prime)^2}$ in fluctuating velocity $\textbf{\textit{u}}^\prime$ obtained in the body-fixed coordinate system by subtracting time-averaged velocity $\overline{\textbf{\textit{u}}}$ from instantaneous velocity $\textbf{\textit{u}}(t)$,
\begin{equation}
\label{TKE:equation}
k = \frac{1}{2}\left[\overline{(u^\prime)^2} + \overline{(v^\prime)^2} + \overline{(w^\prime)^2}\right]. 
\end{equation}

\section{Results}
\label{sec:results}
\subsection{Effect of Sweep Angle on Fluid Forces}

\begin{figure}[hbt!]
\begin{subfigure}{0.49\textwidth}
\centering
\includegraphics[width=\textwidth]{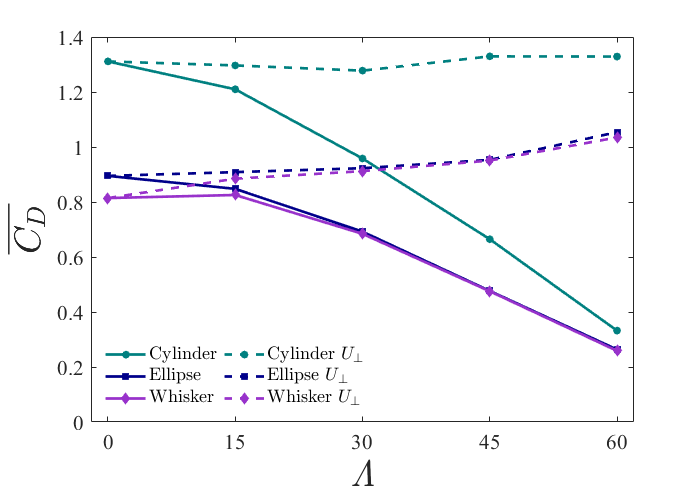}
\caption{Mean coefficient of drag} \label{fig:250a}
\end {subfigure}
\begin{subfigure}{0.49\textwidth}
\centering
\includegraphics[width=\textwidth]{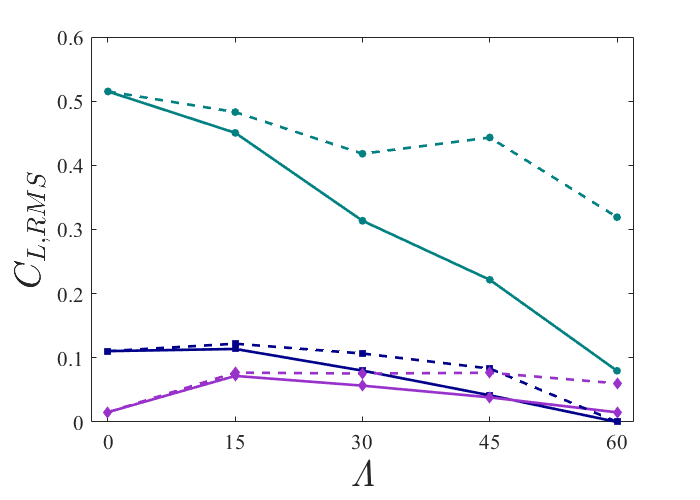}
\caption{RMS lift coefficient} \label{fig:250b}
\end {subfigure}
\caption{Time-averaged drag and root mean square lift coefficient values for the circular cylinder, ellipse, and undulated whisker geometry with respect to sweep angle at $\bm{Re = 250}$. Solid curves are normalized by a constant freestream velocity while dashed lines represent normalizing by the span-perpendicular velocity component in each orientation.}
\label{fig:forcecoef}
\end{figure}

The time-averaged coefficient of drag and root mean square lift coefficient are reported in Figure~\ref{fig:forcecoef} for the three geometries and five sweep angle values at $Re$ = 250. Focusing first on the two smooth geometries of the cylinder and ellipse, there is a clear decrease in forces with respect to increasing sweep angle.  The cylinder's initial drag coefficient of 1.31 decreases to 0.33 when swept at 60 degrees, whereas the streamlined ellipse has $C_D=0.89$ for perpendicular flow and $C_D=0.26$ at $\Lambda=60$ degrees. For smooth bluff bodies, the monotonic reduction in force with increasing sweep is a consequence of decreasing the span-perpendicular velocity component that dominates pressure drag and near-wake forcing. This sweep-induced attenuation provides a comparison against which the effects of whisker undulation-induced flow modification can be evaluated.

At $\mathit{\Lambda} = 0$, the whisker geometry exhibits strong drag reduction with respect to a cylinder ($37.9\%$) and a smaller reduction compared to an ellipse (9.0\%). However, as sweep angle increases, the magnitude of drag forces for the whisker geometry quickly merges with and becomes indistinct from the smooth ellipse at $\mathit{\Lambda} = 30$ and beyond, with both aligning with the circular cylinder at $\mathit{\Lambda} = 60$. Thus, drag coefficient reduction attributable to the undulated surface is only observed when ($\mathit{\Lambda} < 30$).

Observing unsteady lift amplitude in Figure~\ref{fig:250b}, a circular cylinder and ellipse exhibit a large difference at $\mathit{\Lambda} = 0$ with RMS coefficients of 0.51 and 0.11, respectively. As sweep is added, this difference narrows as their forces both decline. This behavior reflects the same sweep-induced reduction observed in drag, where decreasing span-perpendicular flow weakens coherent vortex shedding and associated lift fluctuations for smooth geometries.

Unlike drag, the root mean square lift exhibits stark differences between the whisker and ellipse. For perpendicular flow, the undulated surface drastically reduces $C_{L,RMS}$ compared to a circular cylinder and ellipse alike, with a reduction of 97.0\% and 86.1\%. Unlike the behavior of the smooth geometries, the whisker $C_{L,RMS}$ increases with low sweep angles, rising from 0.02 at $\mathit{\Lambda} = 0$ to 0.07 at $\mathit{\Lambda} = 15$. This increase indicates a weakening of the suppression effects of whisker undulations with the introduction of sweep, as spanwise flow components disrupt the three-dimensional wake interactions responsible for lift reduction. This still maintains a significantly smaller amplitude than an ellipse with a reduction of 36.9\%. The $C_{L,RMS}$ value remains below an ellipse until a sweep angle of $\mathit{\Lambda} = 45$ at which point values for an ellipse and whisker are roughly equivalent. This convergence represents a crossover between two competing effects: the monotonic sweep-induced reduction in unsteady forcing and the diminishing influence of whisker-induced lift suppression with increasing sweep. At this point, there is a notable similarity in forces magnitudes arising from different wake structures between each geometry, and a likely marker of diminishing returns in the effectiveness of the modeled whisker-inspired undulations at or near this sweep angle. To underscore this, at $\mathit{\Lambda} = 60$, the undulated surface $C_{L,RMS}$ is higher than the ellipse though still below the circular cylinder and ellipse model. 

The dashed curves in Figure~\ref{fig:forcecoef} represent the force coefficients normalized by the perpendicular velocity component. Cylinder drag forces collapse onto a nearly flat trend, suggesting that the perpendicular flow component dominates the sweep angle effects for this geometry. Here, $C_{D\perp}$ reflects the approximate validity of the classical independence principle for a cylinder where drag scales with the perpendicular velocity component. The whisker and ellipse, however, show a noticeable upward trend at sweep angles of $\mathit{\Lambda} = 30 - 60$, indicating additional contributions to drag from geometry-specific wake interactions that are over-corrected. This demonstrates a departure from IP predictions, particularly for the whisker geometry's initial increase in drag at 15 degrees which is not accounted for in models normalizing by perpendicular velocity. Lift coefficients for all geometries are more complex, indicating the effect of this renormalization is less significant than for drag forces, suggesting that lift is perhaps more sensitive to the three-dimensional wake dynamics explored in later sections.

\begin{figure}[hbt!]
\begin{subfigure}{0.49\textwidth}
\centering
\includegraphics[width=\textwidth]{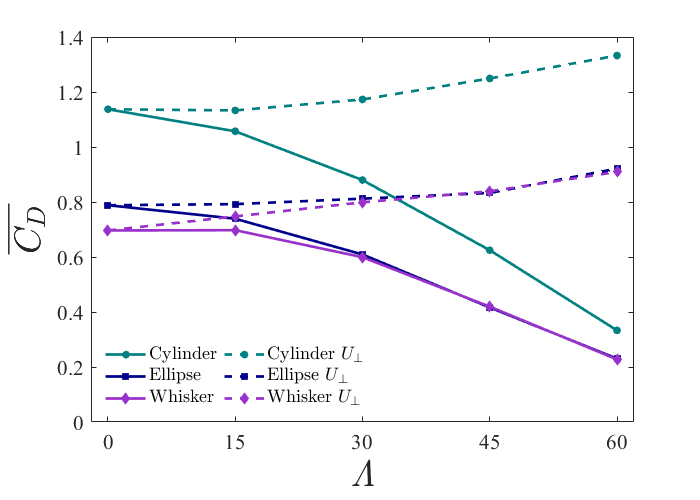}
\caption{Mean coefficient of drag} \label{fig:500a}
\end{subfigure}
\begin{subfigure}{0.49\textwidth}
\centering
\includegraphics[width=\textwidth]{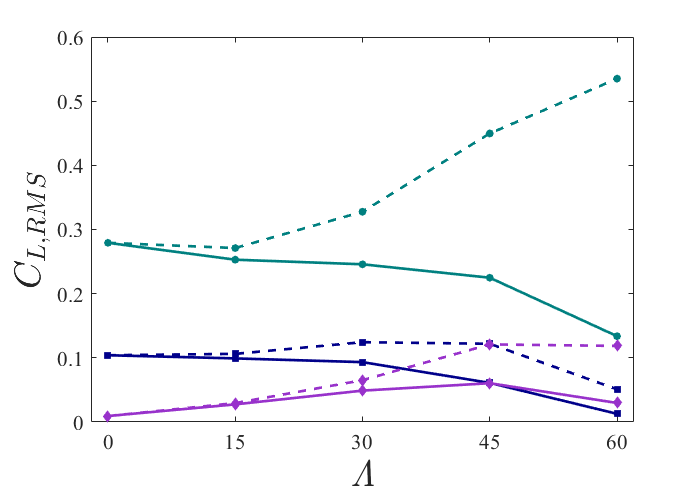}
\caption{RMS lift coefficient} \label{fig:500b}
\end{subfigure}
\caption{Averaged force coefficient values  for the cylinder, ellipse, and whisker geometries with respect to sweep angle at $\bm{Re = 500}$ normalized by a constant freestream velocity and alternatively the span-perpendicular velocity component in each orientation.}
\label{fig:forcecoef500}
\end{figure}

Simulations at Reynolds number 500 exhibit anticipated stronger fluctuations and less orderly wake structures than $Re = 250$. Forces follow similar expected mean trends in sweep-induced responses, and these expectations help distinguish between changes due to Reynolds number and undulation-specific effects. Drag coefficients for all geometries again decrease with increasing sweep angle, however the whisker shows a slightly stronger effective drag suppression with a larger reduction from the ellipse in perpendicular flow (11.4\% versus 9.0\%) with suppression persisting at $\mathit{\Lambda} = 15$. Perpendicular coefficients diverge further from expected collapse of truly two-dimensional, perpendicular velocity scaled flows at this Reynolds number, particularly for larger sweep angles that are similarly over-corrected. The most significant change occurs in lift forces. In addition to the decrease in $C_{L,RMS}$ for all geometries at this Reynolds number, the whisker exhibits an even stronger suppression in perpendicular flow at 90.8\% relative to an ellipse. While whisker RMS lift increases with sweep angle, the magnitude of the suppression remains higher for longer as the whisker demonstrates lower lift forces at $\mathit{\Lambda} = 15$ and $30$ before converging with an ellipse at $\mathit{\Lambda}=45$. This point represents a whisker lift maximum and reflects a similar observation of competing effects identified at $Re = 250$. Here, degradation of whisker-induced suppression mechanisms reaches a balance with sweep-induced attenuation; however, at this higher Reynolds number, the former remained in greater effect over a broader range of sweep angles. The ellipse and whisker force magnitudes again demonstrate an interesting convergence despite differing wakes, then diverge at $\mathit{\Lambda}=60$ similarly to $Re = 250$. Finally, sweep-dependent lift magnitudes are also poorly collapsed by velocity scaling in this regime as flow conditions increasingly violate assumptions of quasi-two-dimensional flow necessary for estimation using IP, as will be shown in later sections.

\begin{figure}[hbt!]
\centering
\begin{subfigure}[tb]{0.48\textwidth}
\centering
\includegraphics[width=5cm,clip=false]{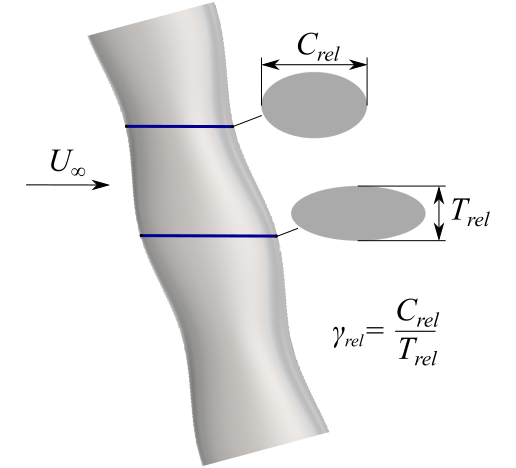}
\caption{Relative chord and thickness at $\bm{\mathit{\Lambda} = 15}$}
\end{subfigure}
\begin{subfigure}[tb]{0.48\textwidth}
\centering
\includegraphics[width=5cm,clip=false]{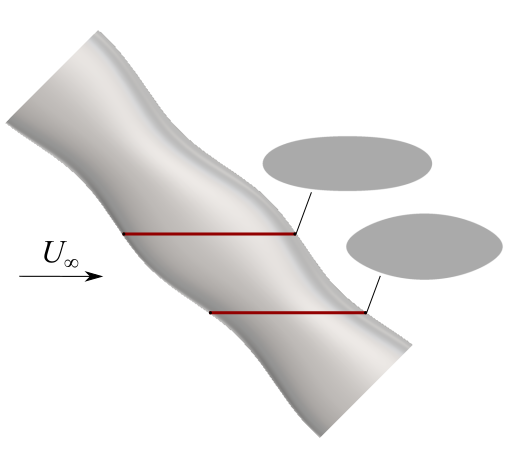}
\caption{Relative chord and thickness at $\bm{\mathit{\Lambda} = 45}$}
\end{subfigure}
\begin{subfigure}[tb]{0.475\textwidth}
\centering
\includegraphics[width=\textwidth]{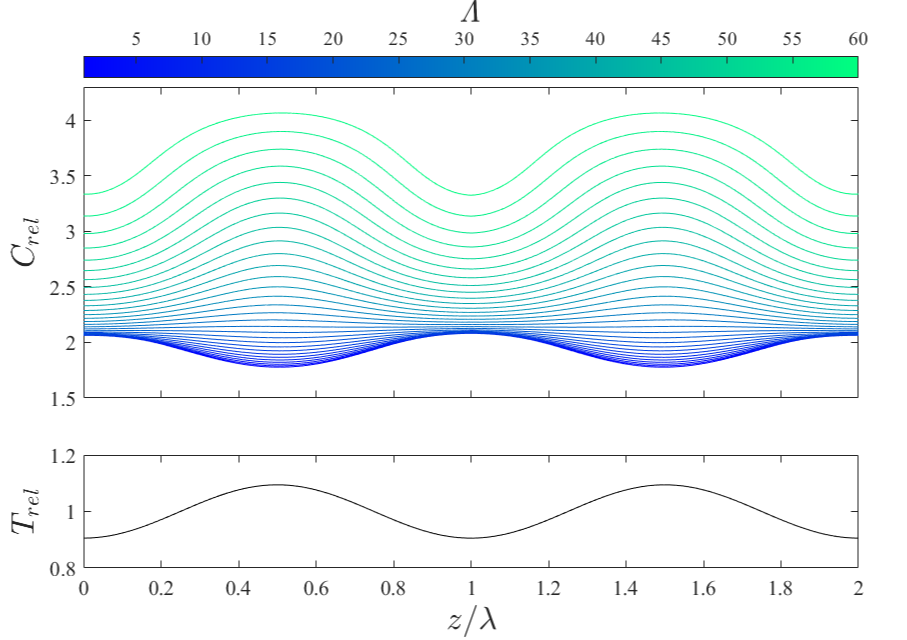}
\caption{Variation in chord length $\bm{C}$ and thickness $\bm{T}$}
\label{fig:relative_chord_thickness}
\end{subfigure}
\begin{subfigure}[tb]{0.475\textwidth}
\centering
\includegraphics[width=\textwidth]{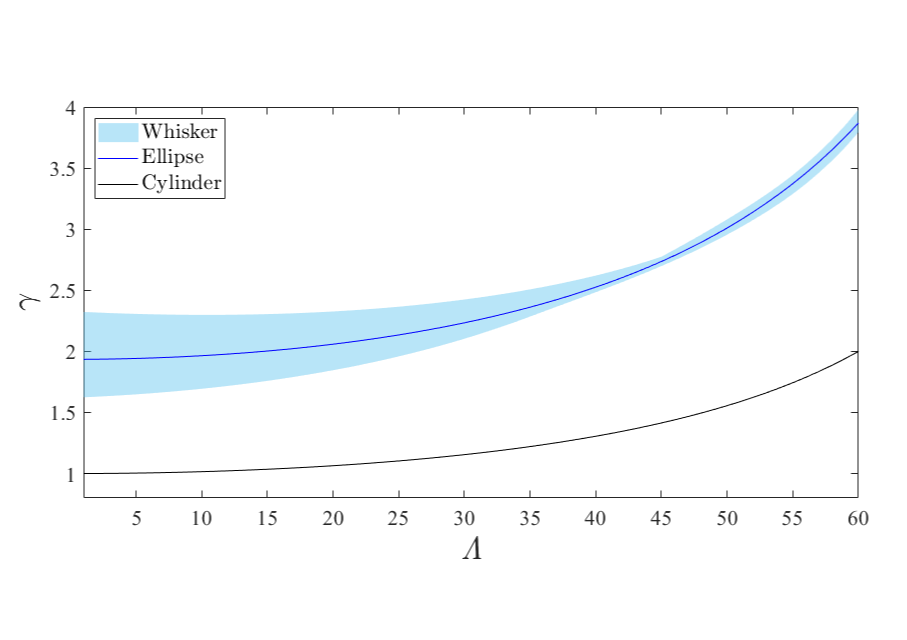}
\caption{Aspect ratio as a function of sweep:  Whisker min/max variation across spanwise positions}
\label{fig:relative_gamma}
\end{subfigure}
\caption{Sweep-dependent spanwise variation in geometry cross section parallel to incoming flow.}
\label{fig:relative_geometry}
\end{figure}

To understand the force observations of the undulated cylinder, Figure~\ref{fig:relative_geometry} describes the interaction between sweep angle and the relative cross section. While possessing the same mean chord length, thickness, and resulting mean aspect ratio as the smooth ellipse geometry, the piece-wise sinusoidal surfaces of the whisker geometry vary the effective value of these parameters with respect to spanwise location and orientation. Figure~\ref{fig:relative_chord_thickness} displays the relative thickness, $T_{rel}$ and chord length $C_{rel}$ as a function of spanwise location ($z$). As shown in the $T_{rel}$ plot, thickness varies with location along the whisker surface, however its magnitude is unaffected by sweep angle. Conversely, relative chord length is dependent on orientation and location. As with a smooth geometry, the overall magnitude of chord length increases with sweep, though the shape of its periodic variation changes. With no sweep angle, the location of maximum chord length aligns with the minimum thickness and vice versa. This relationship gradually changes as sweep angle increases, where at approximately 30 degrees the chord length profile flattens and then reverses its original trend afterwards where maximum chord length and thickness now coincide. These trends are a result of the 180$^{\circ}$ phase difference between thickness and chord undulations as well as the smaller phase difference between leading and trailing undulations in chord length.

Together, these two dimensions create hydrodynamically important effects on the relative aspect ratio of the undulated cylinder, and can explain trends in the forces specific to the surface changes. In Figure~\ref{fig:relative_gamma}, the sweep angle-dependent aspect ratio of the whisker geometry is represented as a vertical range. It is widest at zero sweep angle as the undulation phase difference along the $z$-axis creates cross-sections of thicker, round cylinder-like geometry alternating with regions that have a longer streamlined cross-section. As sweep angle increases, the undulation phase difference relative to sweep is reduced and then reversed, creating an increasingly uniform cross-section across the span. The range of deviation from the mean reaches a minimum at approximately 45 degrees, converging with the ellipse geometry and notably approximating its flow response and resulting forces which converge between the two geometries at this point. 

\subsection{Time-Averaged Streamlines}
Time-averaged streamlines near the whisker surface provide insight into the three-dimensional features of the flow. The resulting streamlines represent the mean surface flow topology (rather than instantaneous vortex structures) and the identified separation features should be interpreted as mean-flow indicators rather than discrete shedding events. Imposing a sweep angle drives a redirection of near-wall flow in the spanwise direction rather than reversing purely in the streamwise direction, and separation does not always correspond to strong reversal of the body-fixed $x$ velocity component. This same effect, however, alters flow characteristics of the undulated geometry in distinct ways from the smooth comparison geometries. 

\begin{figure}[t!]
\begin{subfigure}{\textwidth}
\centering
\includegraphics[width=\textwidth]{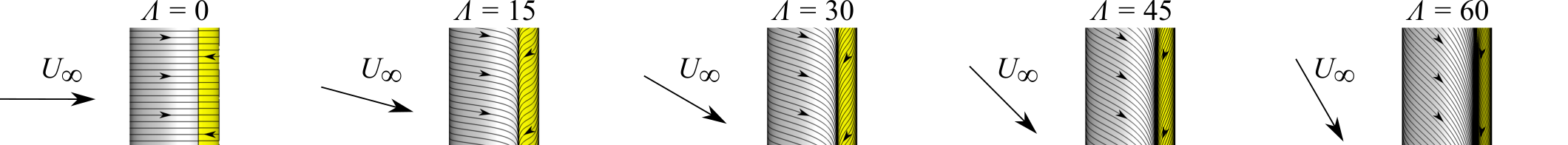}
\caption{$Re = 250$ Smooth Ellipse}
\label{fig:ellipse_SLA}
\end{subfigure}
\begin{subfigure}{\textwidth}
\centering
\includegraphics[width=\textwidth]{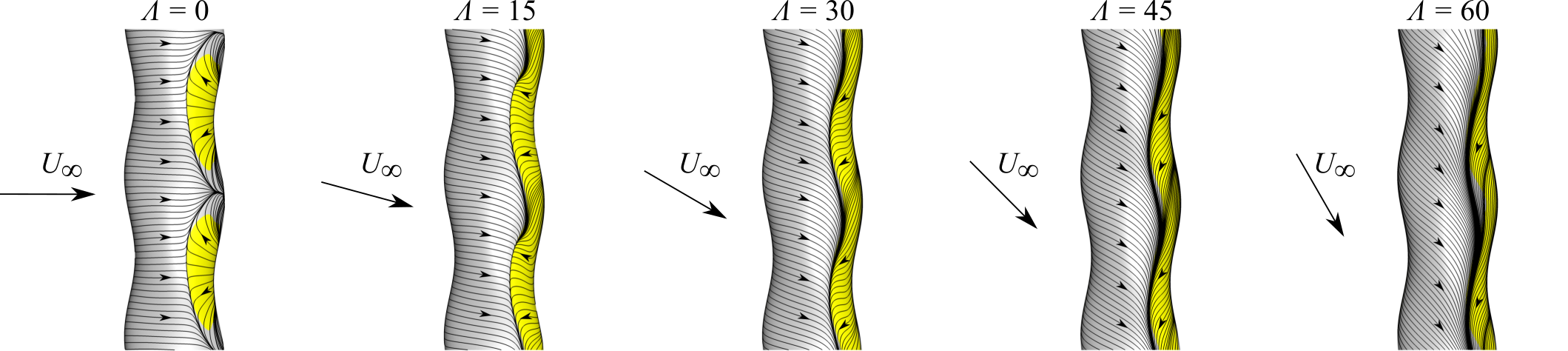}
\caption{$Re = 250$ Undulated Cylinder}
\label{fig:whisker_SLA}
\end{subfigure}
\begin{subfigure}{\textwidth}
\centering
\includegraphics[width=\textwidth]{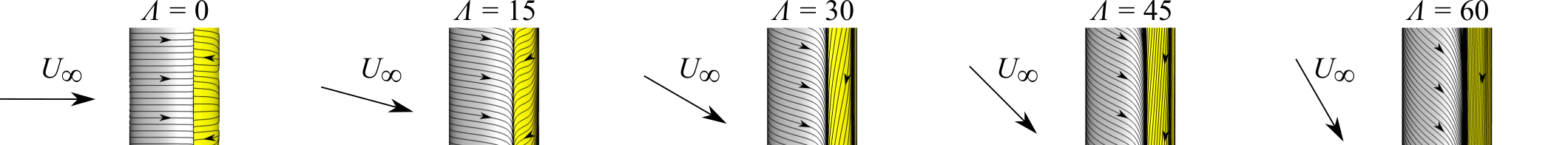}
\caption{$Re = 500$ Smooth Ellipse}
\label{fig:ellipse_SLB}
\end{subfigure}
\begin{subfigure}{\textwidth}
\centering
\includegraphics[width=\textwidth]{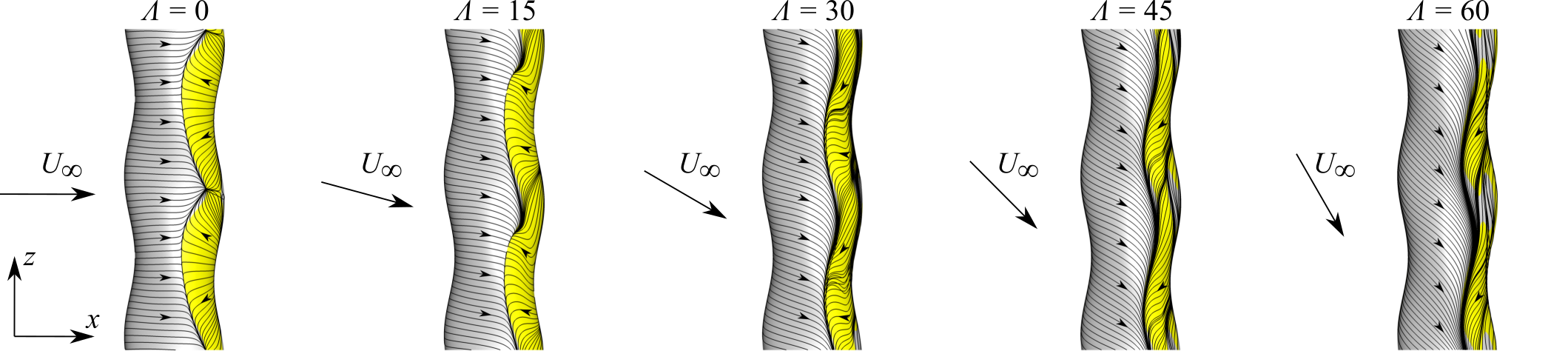}
\caption{$Re = 500$ Undulated Cylinder}
\label{fig:whisker_SLB}
\end{subfigure}
\caption{Surface streamlines of time-averaged velocity $\overline{u}$ with shading representing regions of negative wall shear stress in the body-fixed $x$ direction.}
\label{fig:Streamlines}
\end{figure}

Observing a cropped segment of ellipse geometry at $Re = 250$ in Figure~\ref{fig:ellipse_SLA}, the surfaces demonstrate spanwise-uniform flow and a separation point that is constant across the span. Separation angles indicated in Table~\ref{tab:separationtable} for an ellipse show a small monotonic increase with sweep until $\mathit{\Lambda}=60$ where the flow transitions to a different shedding behavior. Shaded areas correspond to regions where wall shear stress is reversed in the body-fixed $x$ direction as a reverse flow indicator, and generally coincide with the separation points indicated by the mean streamlines. For the whisker geometry in Figure~\ref{fig:whisker_SLA}, streamlines demonstrate the increasing spanwise flow with sweep angle. For the perpendicular flow orientation, streamlines are redirected around areas of increased thickness and converge in regions of increased chord length. Trailing edge streamlines showing reverse flow also highlight a unique separation line that is localized to each wavelength of the whisker surface and creates distinct shedding regions divided in the vicinity of the streamwise undulation peak. These distinct separation cells are also visible in the disconnected regions of reverse wall shear stress, a unique phenomenon for this geometry noticeable later in the shedding phase of instantaneous flow structures. 

With a sweep angle, the spanwise flow initiated by geometry undulations is diluted but not eliminated. At 15 degrees, the separation line that delineated two distinct wake regions has disappeared and is replaced by a continuous separation region, but the exact separation angle still varies across the span within a range noted in Table~\ref{tab:separationtable}. This feature is gradually diminished as sweep angle increases. For the swept 30, 45, and 60 degree configurations, the separation angles increase while the range between minimum and maximum values decrease, as spanwise variation weakens and settles to a function of the spanwise relative chord length. 

This progressive loss of strong, localized separation associated with undulations coincides with the observed convergence of lift fluctuations toward ellipse-like behavior. Also, at $\mathit{\Lambda}=60$, shaded reverse wall shear stress in the $x$-direction no longer entirely coincides with converging streamline-indicated separation, demonstrating a departure from typical shedding behavior due to three-dimensional spanwise flow at the extreme angles. 

Time-averaged streamline flow topology trends are largely similar for an ellipse at $Re = 500$ as demonstrated in Figure~\ref{fig:ellipse_SLB} with an anticipated separation angle reduction compared to $Re = 250$. However, Table~\ref{tab:separationtable} indicates separation angle across sweep does not increase monotonically owing to increasingly three-dimensional instability effects. Unlike $Re = 250$, the seal whisker at $\mathit{\Lambda}=0$ still experiences significant spanwise flow near the locations of maximum chord length. However, entirely distinct shedding cells are now absent which is consistent with the reduced spanwise segmentation of vortex structures shown in later instantaneous wake visualizations. The remaining whisker sweep angles behave in similar fashion to their prior $Re = 250$ counterparts representing similar geometric effects despite the transition towards greater turbulence and instability.

\begin{table}[hbt!]
\caption{Flow separation angles, $\bm{\theta_{sep}}$, measured from the leading stagnation point to the zero crossing of the body-fixed $\bm{x}$-component of wall shear stress. For the whisker geometry, separation spans a band along the model where minimum and maximum values indicate the range which occurs at varying $\bm{z}$ locations along the span.}
\centering
\begin{minipage}[c]{0.6\textwidth} 
\centering
\begin{tabular}{|c|c|c|c|c|c|c|}
\hline
 & \multicolumn{3}{c|}{$Re = 250$} & \multicolumn{3}{c|}{$Re = 500$} \\
\cline{2-7}
$\mathit{\Lambda}$ & \makecell{Ellipse\\($^\circ$)} & \makecell{Whisker\\min. ($^\circ$)} & \makecell{Whisker\\max. ($^\circ$)} & \makecell{Ellipse\\($^\circ$)} & \makecell{Whisker\\min. ($^\circ$)} & \makecell{Whisker\\max. ($^\circ$)} \\
\hline \hline
0  & 140.6 & 116.5 & ---   & 132.1 & 106.2 & 159.2 \\
\rowcolor{gray!25}
15 & 141.4 & 123.8 & 154.4 & 132.2 & 112.3 & 149.0 \\
30 & 142.1 & 127.0 & 152.2 & 126.4 & 116.4 & 143.8 \\
\rowcolor{gray!25}
45 & 143.8 & 128.5 & 153.3 & 127.4 & 118.5 & 147.4 \\
60 & 147.9 & 130.0 & 161.4 & 134.6 & 120.0 & 171.0 \\
\hline
\end{tabular}
\end{minipage}%
\hspace{0.02\textwidth}
\begin{minipage}[c]{0.25\textwidth} 
\centering
\includegraphics[width=0.8\textwidth]{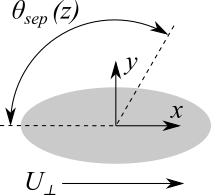}
\end{minipage}
\label{tab:separationtable}
\end{table}

\subsection{Vortex Wake Effects}
\begin{figure}[hbt!]
\begin{subfigure}{\textwidth}
\centering
\label{fig:250Q00}
\includegraphics[width=\textwidth]{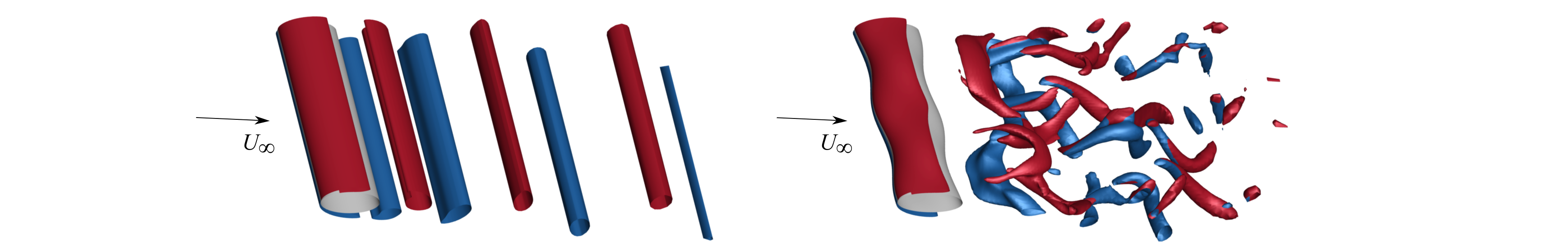}
\caption{$\mathit{\Lambda} = 0, Q = 0.4$} 
\end{subfigure}
\begin{subfigure}{\textwidth}
\centering
\label{fig:250Q15}
\includegraphics[width=\textwidth]{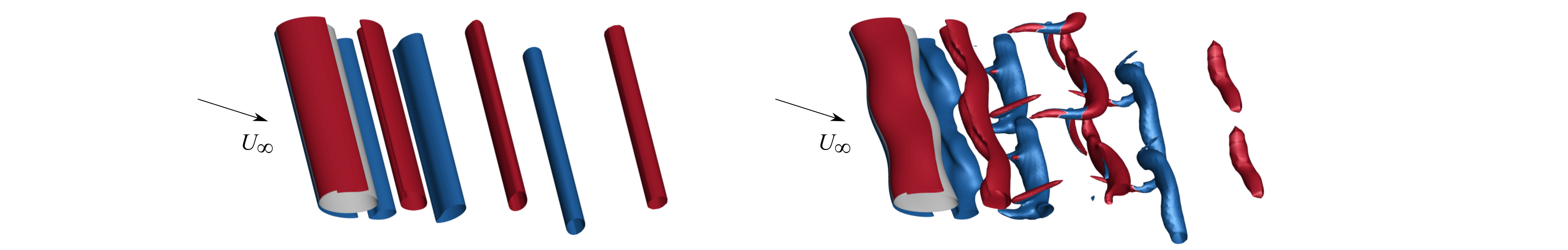}
\caption{$\mathit{\Lambda} = 15, Q = 0.4$} 
\end{subfigure}
\begin{subfigure}{\textwidth}
\centering
\label{fig:250Q30}
\includegraphics[width=\textwidth]{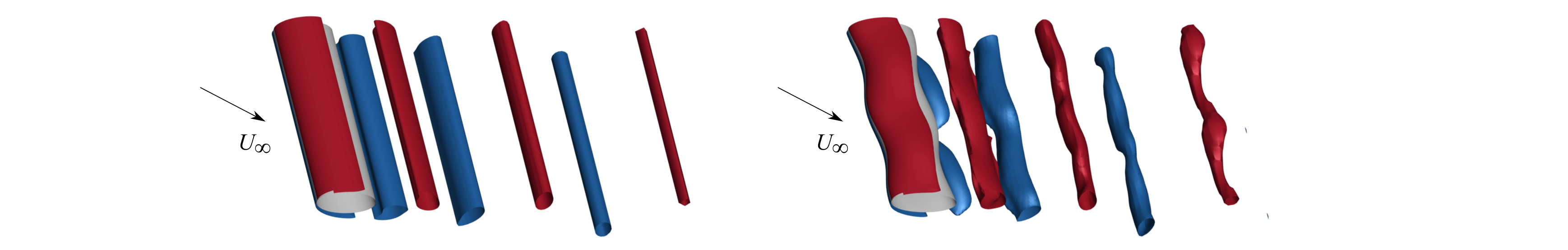}
\caption{$\mathit{\Lambda} = 30, Q = 0.32$} 
\end{subfigure}
\begin{subfigure}{\textwidth}
\centering
\label{fig:250Q45}
\includegraphics[width=\textwidth]{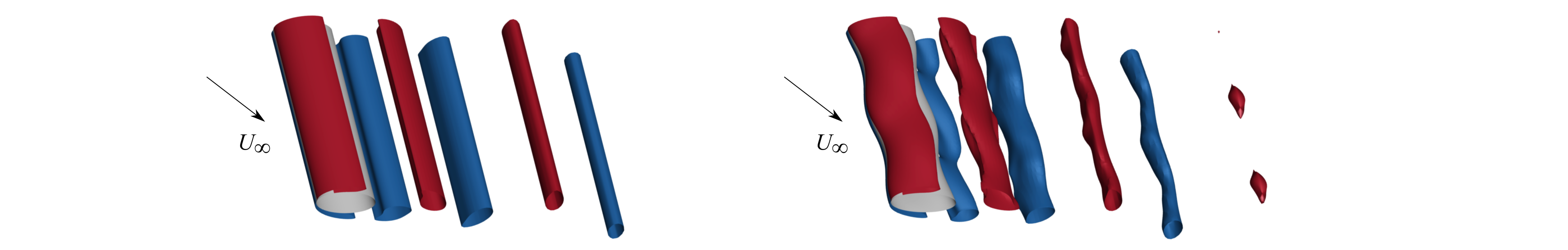}
\caption{$\mathit{\Lambda} = 45, Q = 0.16$}
\end{subfigure}
\begin{subfigure}{\textwidth}
\centering
\label{fig:250Q60}
\includegraphics[width=\textwidth]{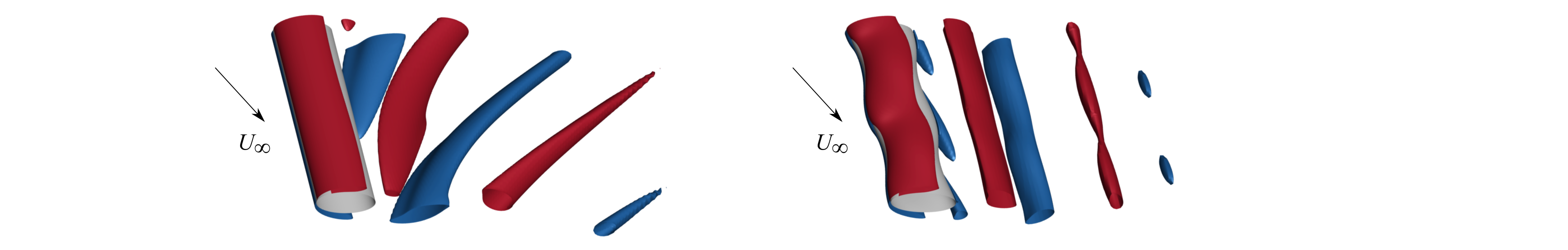}
\caption{$\mathit{\Lambda} = 60, Q = 0.08$} 
\end{subfigure}
\caption{Flow structures for $\bm{Re = 250}$ visualized by isosurfaces of $\bm{Q}$-criterion colored by the sign of $\bm{z}$-vorticity. As sweep angle increases the fluid rotation strength on the axis parallel to the span ($\bm{\omega_z}$) diminishes, thus the value of $\bm{Q}$ visualized is lower as sweep increases.}
\label{fig:Q250split}
\end{figure}

The flow structures created by the surface undulations are markedly three-dimensional in comparison to the smooth geometry, strongly influencing mixing and dissipation of energy in the wake. Figure~\ref{fig:Q250split} shows isosurfaces of $Q$-criterion for the smooth ellipse and undulated cylinder at Reynolds number 250. The wake visualizations shown are representative snapshots selected at comparable points in the shedding cycle.

The salient features of the ellipse at  $\mathit{\Lambda} \leq 45$ are the spanwise symmetry in shed vortices, decreasing shedding frequency, and two-dimensional behavior. At the largest sweep angle of $\mathit{\Lambda} = 60$, vortices are shed at a shallower angle, almost perpendicular to the sweep angle, a pattern noted for similar flows in prior literature \cite{lucor_effects_2003}. 

In contrast, the seal whisker wake at $\mathit{\Lambda} = 0$ consists of two distinct shedding regions that correspond to adjacent wavelengths of the whisker surface. This breakup of vortices along the span is a result of spanwise momentum transport over the undulated surface and leads to a noticeable increase in streamwise vorticity \cite{lyons_wavelength_2022}. Also of note is the alternating phase pattern with which adjacent wavelengths exhibit shedding, a pattern that disappears with the addition of sweep.
For $\mathit{\Lambda} = 15$ small streamwise vortex structures still exist but are much smaller. At $\mathit{\Lambda} = 30$ and $45$, shed vortices are similar to those of the smooth ellipse with residual wavelength-scale separation in the near wake merging into more span-symmetric roller vortices downstream. While visible artifacts of the whisker undulations remain, the primary mechanisms governing lift become increasingly dominated by flow components normal to the projected surface, leading to converging RMS lift coefficient values between the whisker and ellipse geometries. This convergence occurs despite qualitative differences in instantaneous wake structures, illustrating how similar global force statistics shown in Figure~\ref{fig:250a} can arise from distinct local flow dynamics. As sweep increases, the decreasing perpendicular flow component over an ellipse results in weaker roller vortex structures, whereas the whisker’s force-suppression capability, effective at disrupting spanwise coherence at low sweep, becomes increasingly outweighed by sweep-induced spanwise transport that favors re-merging of wake structures and ellipse-like force behavior. 
At $\mathit{\Lambda} = 60$, the whisker geometry is observed to delay the ellipse behavior of peeling off vortex structures at a shallow angle and instead maintains the roller-type structures for a larger sweep. From the perspective of forces, this delayed transition in shedding behavior is less effective at lift suppression compared to the ellipse.

\begin{figure}[hbt!]
\begin{subfigure}{\textwidth}
\centering
\label{fig:500Q00}
\includegraphics[width=\textwidth]{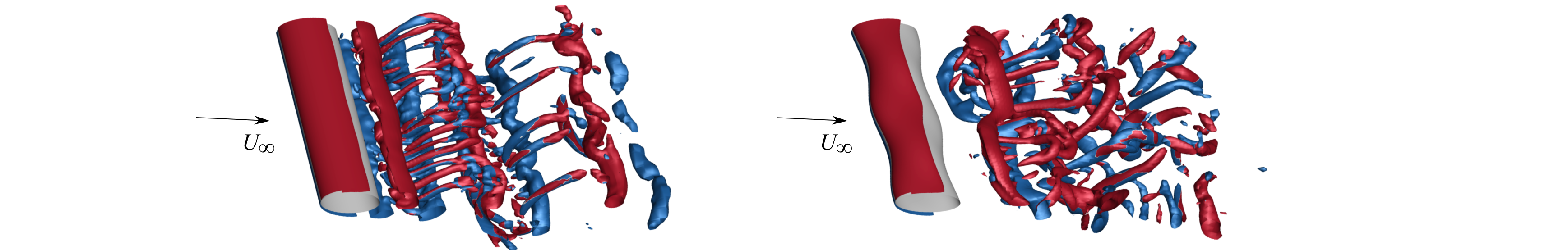}
\caption{$\mathit{\Lambda} = 0, Q = 0.6$}
\end{subfigure}
\begin{subfigure}{\textwidth}
\centering
\label{fig:500Q15}
\includegraphics[width=\textwidth]{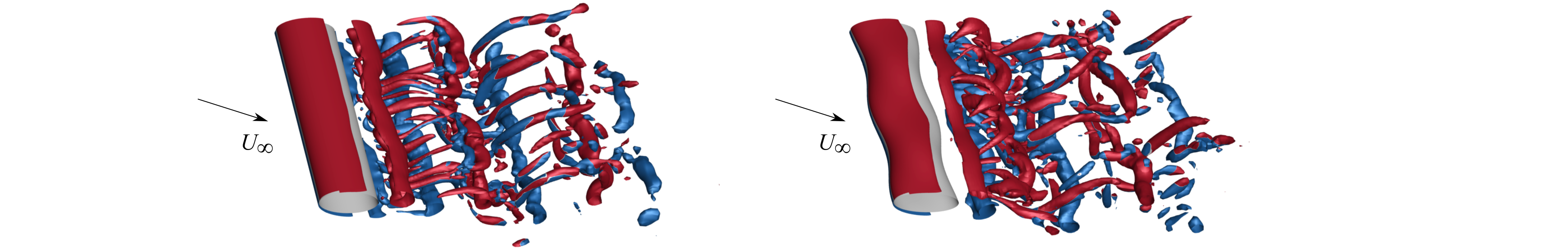}
\caption{$\mathit{\Lambda} = 15, Q = 0.6$} 
\end{subfigure}
\begin{subfigure}{\textwidth}
\centering 
\label{fig:500Q30}
\includegraphics[width=\textwidth]{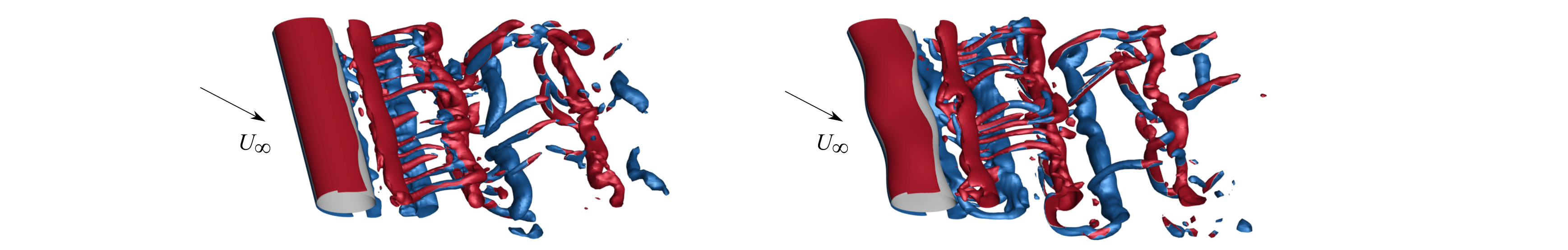}
\caption{$\mathit{\Lambda} = 30, Q = 0.4$}
\end{subfigure}
\begin{subfigure}{\textwidth}
\centering
\label{fig:500Q45}
\includegraphics[width=\textwidth]{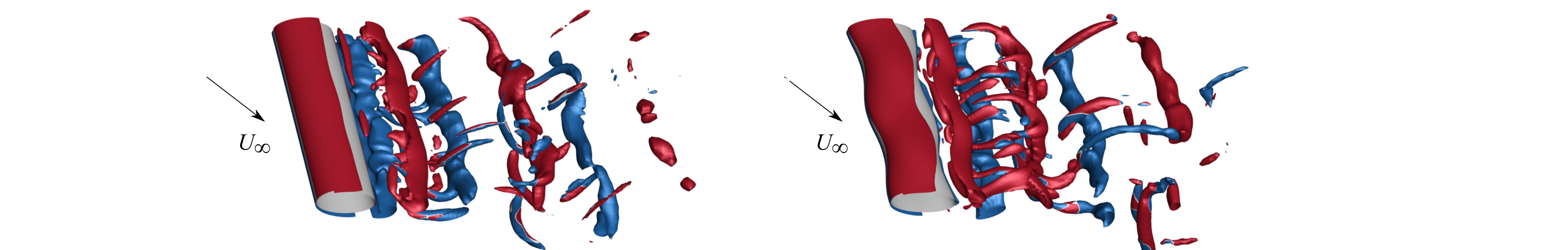}
\caption{$\mathit{\Lambda} = 45, Q = 0.32$} 
\end{subfigure}
\begin{subfigure}{\textwidth}
\centering
\label{fig:500Q60}
\includegraphics[width=\textwidth]{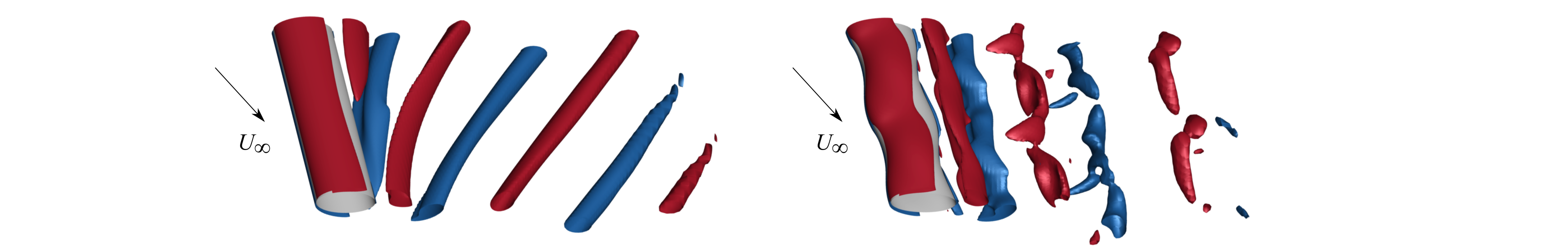}
\caption{$\mathit{\Lambda} = 60, Q = 0.16$} 
\end{subfigure}
\caption{Isosurfaces of $\bm{Q}$-criterion colored by the sign of $\bm{z}$-vorticity for $\bm{Re=500}$. Due to increased vortex strength, higher $\bm{Q}$ values are displayed compared to those reported for $\bm{Re=250}$.}
\label{fig:Q500split}
\end{figure}

The wake structures for $Re = 500$ are shown in Figure~\ref{fig:Q500split}, demonstrating increased local turbulence with finer and more complex vortex structures. Additionally, spanwise instabilities characteristic of the flow regime are prevalent, particularly visible for the ellipse in the form of many streamwise finger-like structures along the span. However, many of the main features and trends observed at $Re=250$ are still present. 
At $\mathit{\Lambda} \leq 15$, the smooth geometry forms stronger and more compact vortices close to the trailing edge compared to the whisker, reflective of the higher RMS lift values. In contrast, the undulated cylinder sheds vortices further from the model surface. For ${\Lambda} = 0$ there is a notable phase difference in the shedding of horseshoe-type vortex structures from adjacent wavelengths, similar to $Re=250$. Likewise, when ${\Lambda} > 0$, there is increasing spanwise symmetry similar to the lower Reynolds number observations, causing the whisker geometry RMS lift to gradually approach that of the ellipse with the $\mathit{\Lambda}$ = 60 whisker geometry once again delaying the transition of the flow and maintaining the roller-like structures, reflecting in the larger RMS lift compared to the ellipse.

It is worth noting that other studies, e.g. \cite{menon_spanwise_2022}, have extended the use of the $Q$-criterion by incorporating both positive and negative $Q$ values in combination with force partitioning methods to enable a deeper dissection of the interplay between vortex dynamics and force generation. These methods highlight the potential for future work in the area of whisker-inspired geometry, particularly in relation to the forces observed as a result three-dimensional wake of this complex geometry.

\begin{figure}[hbt!]
\begin{subfigure}{\textwidth}
\centering
\includegraphics[width=\textwidth]{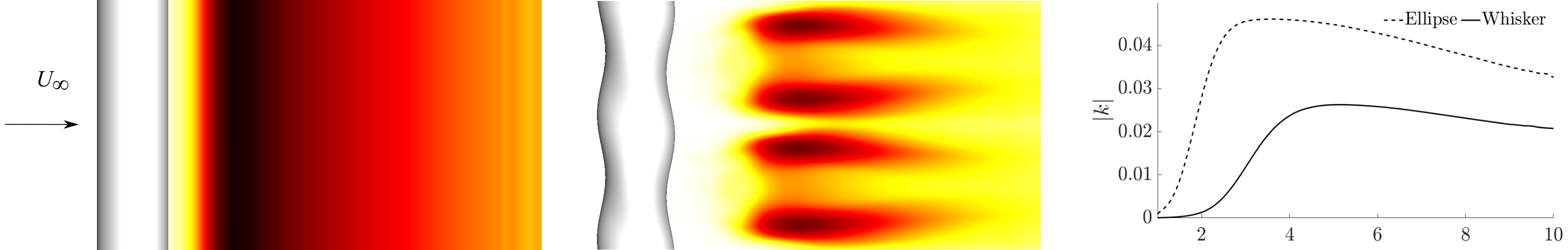}
\caption{$\mathit{\Lambda} = 0$} 
\label{fig:TKE00}
\end{subfigure}
\begin{subfigure}{\textwidth}
\centering 
\includegraphics[width=\textwidth]{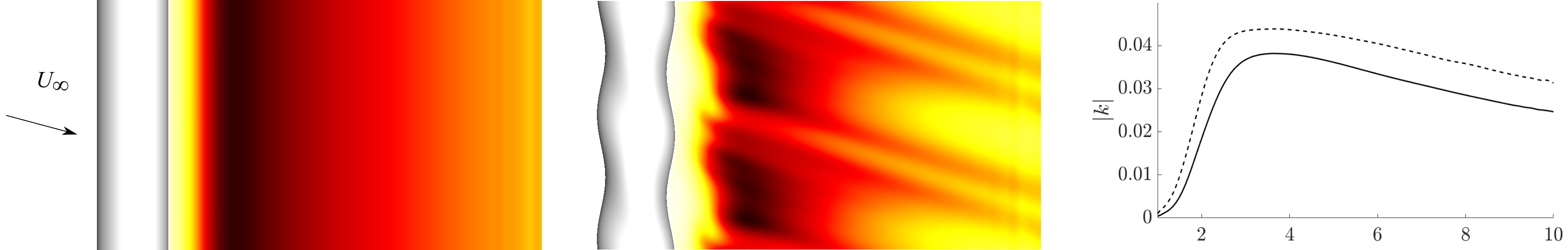}
\caption{$\mathit{\Lambda} = 15$}
\label{fig:TKE15}
\end{subfigure}
\begin{subfigure}{\textwidth}
\centering
\includegraphics[width=\textwidth]{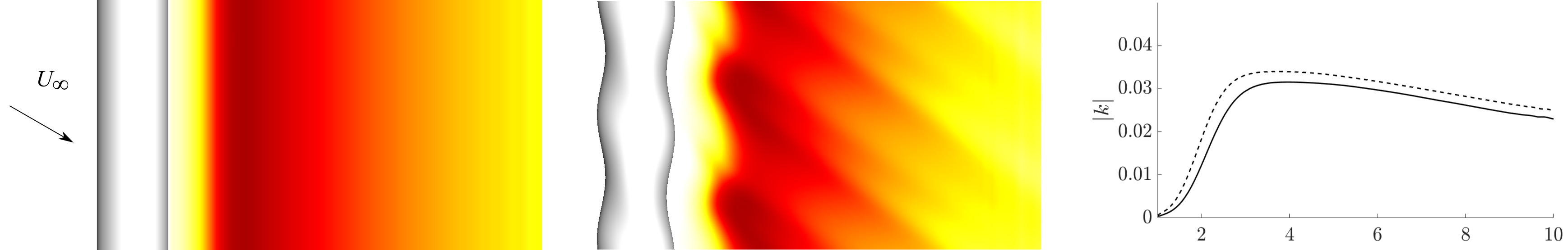}
\caption{$\mathit{\Lambda} = 30$} 
\label{fig:TKE30}
\end{subfigure}
\begin{subfigure}{\textwidth}
\centering
\includegraphics[width=\textwidth]{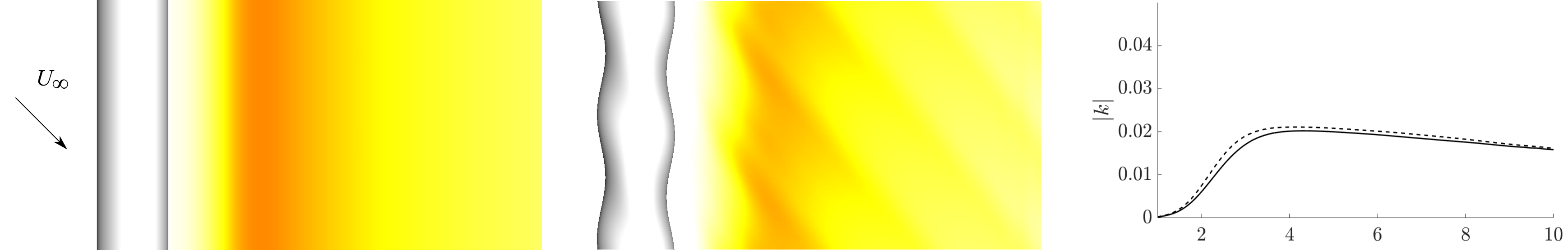}
\caption{$\mathit{\Lambda} = 45$}
\label{fig:TKE45}
\end{subfigure}
\begin{subfigure}{\textwidth}
\centering 
\includegraphics[width=\textwidth]{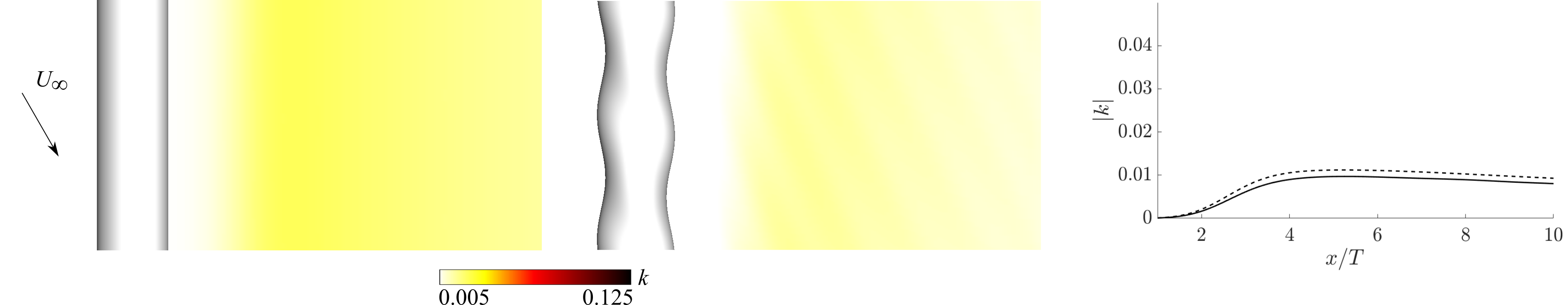}
\caption{$\mathit{\Lambda} = 60$}
\label{fig:TKE60}
\end{subfigure}
\caption{Contours of $\bm{Re = 250}$ TKE in the wake accompanied by plots of span-averaged TKE.}
\label{fig:TKEsplit}
\end{figure}

Turbulent kinetic energy in the wake can be correlated with observed vortex structures, as shown by Figure~\ref{fig:TKEsplit} for $Re = 250$. As previously observed at $\mathit{\Lambda} = 0$ \cite{lyons_wavelength_2022}, the largest values of TKE occur in periodic spanwise regions that align with the streamwise vortex structures visible in Figure~\ref{fig:Q250split} and, compared to a smooth ellipse, exhibit a lower peak magnitude and form further downstream in the wake. 
At $\mathit{\Lambda} = 15$, the influence of altered streamwise vortex structures is still visible in contours of TKE where their interference with adjacent structures creates long pairs of streamwise mixing regions.

There is also a strong increase in TKE magnitude, correlating the increase in unsteady lift, when sweep is introduced at 15 degrees.  Figure~\ref{fig:TKEsplit}, column three, demonstrates this observation through span-averaged TKE as a function of $x/T$.  Figure~\ref{fig:TKE00} shows a large decrease in TKE magnitude for the whisker geometry. As sweep is introduced TKE magnitude decreases for the ellipse but initially jumps for the whisker geometry, mirroring the behavior of the force coefficients. Beyond this point, the magnitude of wake TKE decreases steadily with sweep angle to minimum values at $\mathit{\Lambda} = 60$. The downstream $x$ location at which TKE peak magnitudes occur is also strongly influenced by geometry prior to sweep and then becomes relatively uniform with increasing angle. 

For $\mathit{\Lambda}= 30$ and beyond, a periodic TKE pattern is visible in the wake  however the magnitude with respect to spanwise location becomes increasingly uniform. For $\mathit{\Lambda} = 45$, spanwise momentum transport is now dominated by the freestream spanwise component instead of surface undulations, and the resulting wake increasingly mirrors that of an ellipse. At $\mathit{\Lambda} = 60$, as is shown by flow structures in Figure~\ref{fig:Q250split}, the presence of undulations appears to delay the transition of vortex shedding to a shallow angle regime with a noticeable mismatch from the sweep angle that is also reflected in contours of TKE.

\begin{figure}[hbt!]
\begin{subfigure}{\textwidth}
\centering
\includegraphics[width=\textwidth]{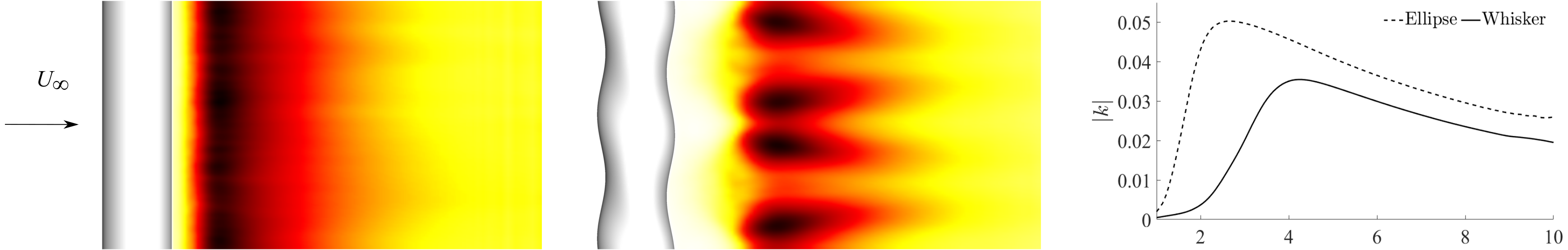}
\caption{$\mathit{\Lambda} = 0$} 
\label{fig:TKE00500}
\end{subfigure}
\begin{subfigure}{\textwidth}
\centering 
\includegraphics[width=\textwidth]{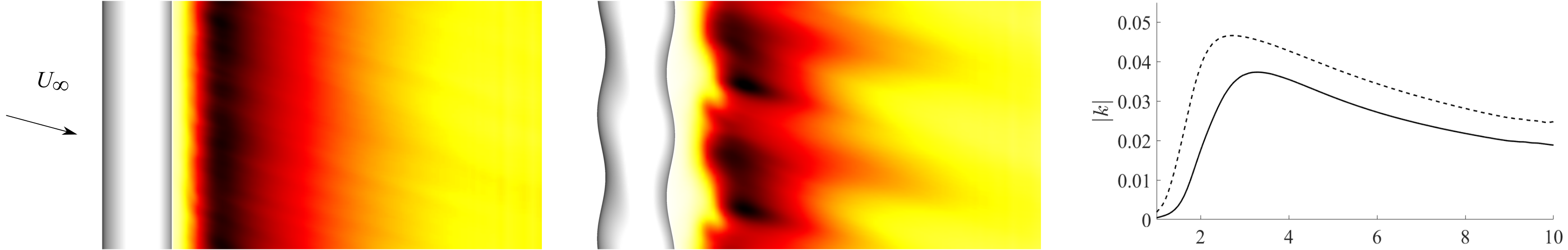}
\caption{$\mathit{\Lambda} = 15$}
\label{fig:TKE15500}
\end{subfigure}
\begin{subfigure}{\textwidth}
\centering
\includegraphics[width=\textwidth]{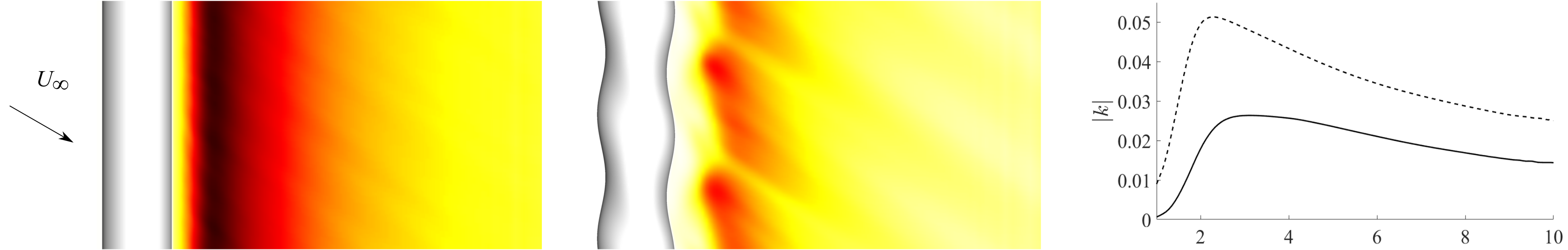}
\caption{$\mathit{\Lambda} = 30$} 
\label{fig:TKE30500}
\end{subfigure}
\begin{subfigure}{\textwidth}
\centering
\includegraphics[width=\textwidth]{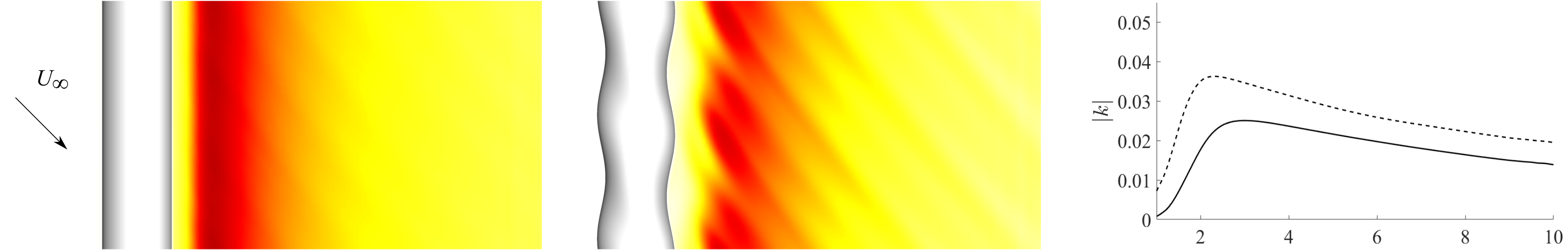}
\caption{$\mathit{\Lambda} = 45$}
\label{fig:TKE45500}
\end{subfigure}
\begin{subfigure}{\textwidth}
\centering 
\includegraphics[width=\textwidth]{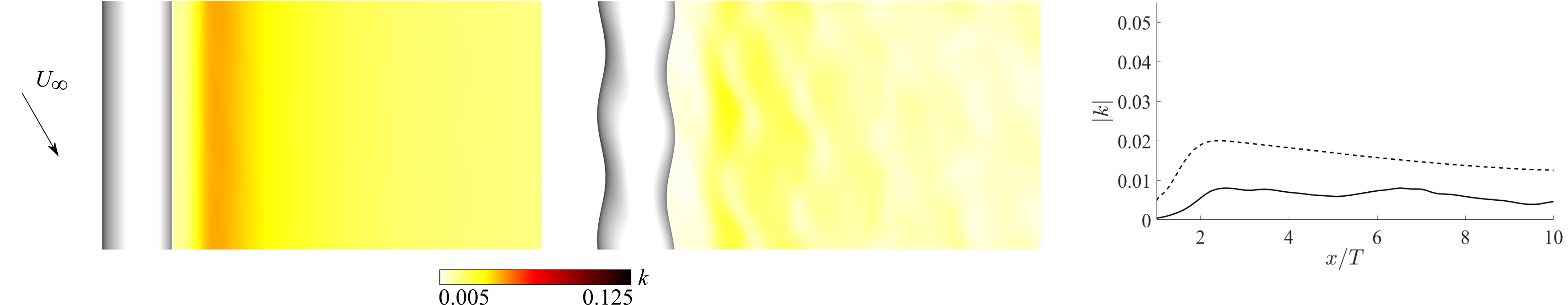}
\caption{$\mathit{\Lambda} = 60$}
\label{fig:TKE60500}
\end{subfigure}
\caption{Contours of $\bm{Re = 500}$ TKE in the wake accompanied by plots of span-averaged TKE.}
\label{fig:TKEsplit500}
\end{figure}

At $Re = 500$, shown in Figure~\ref{fig:TKEsplit500}, flow is more unstable and prone to three-dimensional breakdown, yet the overall spatial reorganization of TKE remains consistent with trends observed at $Re = 250$. A smooth ellipse demonstrates an anticipated increase in peak intensity across all contours. The seal whisker demonstrates a similar periodic spanwise pattern at $\mathit{\Lambda = 0}$ with a concentration of TKE in paired structures in the wake of each wavelength. With added sweep, this pattern is progressively disrupted, though periodic influence from the surface undulations remains evident. 
While sweep continues to redistribute TKE in the wake and reduce spanwise variation, an important Reynolds-number effect emerges. Span-averaged TKE shows that the difference between whisker and ellipse geometries not only persists, but becomes significantly larger than at $Re = 250$, particularly for $\mathit{\Lambda \geq 30}$. In this high-sweep regime, force suppression was observed to diminish, yet the whisker maintains substantially lower wake TKE than the ellipse. This indicates that at higher Reynolds number, surface undulations become increasingly effective at redistributing turbulent energy into more three-dimensional convecting structures. Thus, even where force behavior converges, the whisker continues to alter wake turbulence intensity and organization relative to an ellipse.

\subsection{Shedding Frequency Trends}

\begin{figure}[hbt!]
\centering
\begin{subfigure}{0.49\textwidth}
\includegraphics[width=\textwidth]{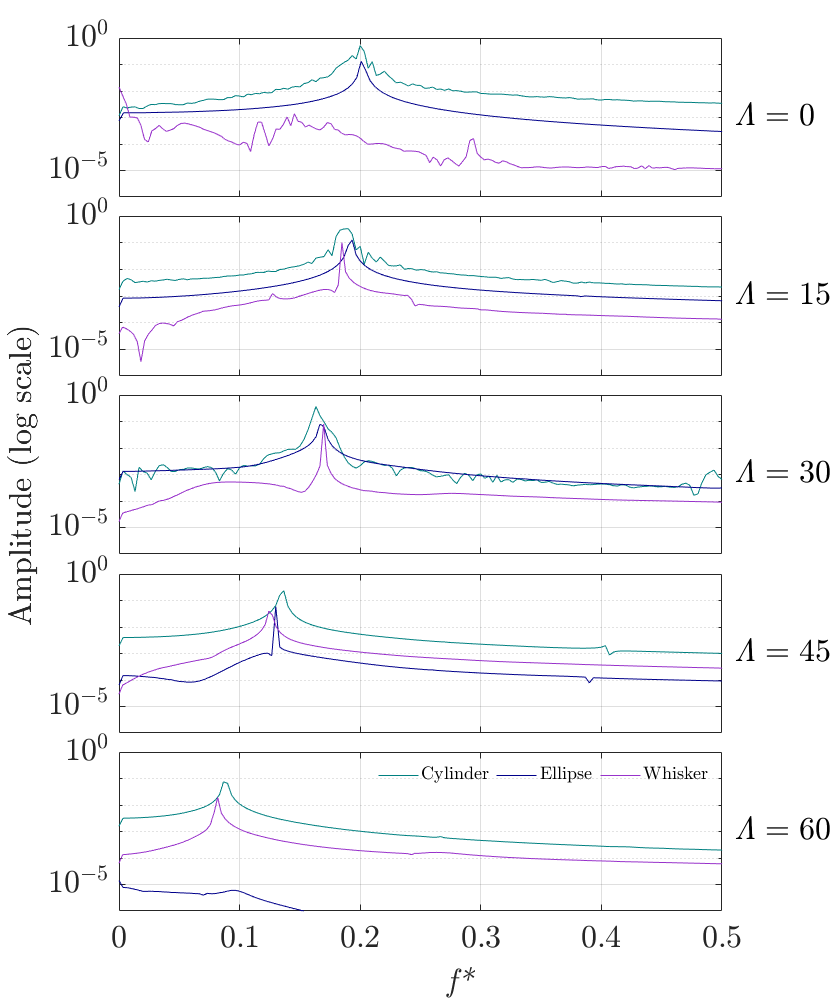}
\caption{$Re = 250$} 
\label{fig:Spectra250}
\end{subfigure}
\begin{subfigure}{0.49\textwidth}
\centering 
\includegraphics[width=\textwidth]{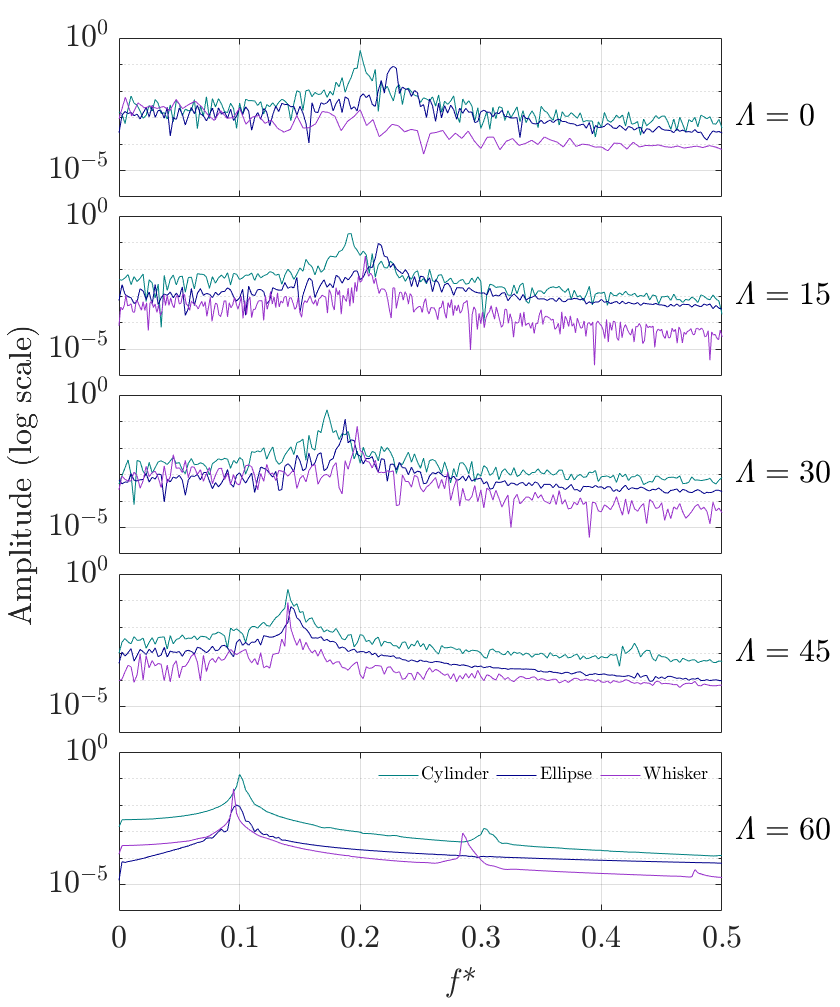}
\caption{$Re = 500$}
\label{fig:Spectra500}
\end{subfigure}
\caption{Spectrum of unsteady lift forcing for all geometries and sweep angles at Reynolds number 250 and 500.}
\label{fig:frequency}
\end{figure}

The differences in the shedding pattern shown by flow structures have the strongest influence on the shedding frequency spectrum of each geometry at smaller sweep. Spectra for the cylinder, ellipse, and whisker across all sweep angles are shown in Figure~\ref{fig:frequency}. As noted in prior work \cite{dunt_frequency_2024}, the $\mathit{\Lambda}=0$ whisker exhibits spanwise phase-shifted shedding with lower spectral magnitude and reduced peak frequency relative to smooth geometries. With the addition of sweep, a coherent shedding peak reemerges and the whisker shedding frequency converges toward the cylinder and ellipse values at $\mathit{\Lambda}=15$ and $30$, though amplitudes remain lower, consistent with reduced RMS lift. At $\mathit{\Lambda}=45$, the whisker spectrum surpasses the ellipse, matching the previously identified break-even point in RMS lift while arising from distinct wake dynamics. At this large sweep angle, the force-suppression capability of the whisker is diminished, and the dominant shedding frequency across geometries appears to be governed primarily by the effective bluff-body dimensions which are similar for an ellipse and whisker but distinct from the cylinder. Then, at $\mathit{\Lambda}=60$, the whisker is observed to delay the ellipse's transition to a new shedding regime, retaining stronger periodic lift fluctuations while the ellipse exhibits almost entirely diminished oscillations. Overall, sweep angle causes a gradual realignment of whisker shedding and ellipse-like behavior before later demonstrating the key divergence between the two geometries for the largest sweep angle.

At $Re = 500$, the shedding spectra show greater noise form the more transitioned wake, but key distinctions remain. The difference between peak frequencies across all geometries is initially greater, but all once again converge to similar values at $\mathit{\Lambda}=45$. Notably, the largest sweep angle $\mathit{\Lambda}=60$ demonstrates smoother spectra for all geometries representing a delayed introduction of three-dimensional flow effects and turbulent transition. Additionally, the smooth ellipse at this angle now exhibits a coherent peak as opposed to its diminished periodic shedding behavior at lower Reynolds number.

\section{Conclusion}
\label{sec:conclusion}
Numerical simulations of swept flow over a whisker-inspired geometry are investigated at Reynolds numbers of 250 and 500. A comparison between the whisker geometry and a smooth circular cylinder reveals a significant reduction in unsteady lift and drag forces across the entire range of tested sweep angles. Comparison with a smooth ellipse of equivalent aspect ratio reveals a reduction of forces for the range of flow angles of $\mathit{\Lambda}$ = 0 to 45 and a slight increase in RMS lift at $\mathit{\Lambda} = 60$. The impact of surface undulations is strongest in flow perpendicular to the model span ($\mathit{\Lambda} = 0$) with a reduction in RMS lift force by 86.1\% and 90.8\% from that of a smooth ellipse for $Re=250$ and $Re=500$, respectively. In terms of drag forces, there is a reduction of 9.0\% and 11.4\% as sweep angle increases forces are  reduced, by a smaller magnitude, up to a sweep angle of $\mathit{\Lambda} = 45$ where the ellipse and whisker converge to similar values before ultimately diverging with a larger whisker lift force at $\mathit{\Lambda} = 60$. 

The forces on the whisker are influenced by the three-dimensional flow caused by the spanwise-varied cross-section. This has a complex evolution with respect to increasing sweep angle since the relative cross-section is a function of $\mathit{\Lambda}$. With larger sweep angle, the whisker undulations increasingly approximate the relative aspect ratio of an ellipse, and thus the strong spanwise flow effects generated by surface undulations are observed to gradually give way to ellipse-like shedding behavior. 

While perpendicular flow has distinct shedding regions along the span, large sweep angles merge these shedding regions and approximate the flow structures of a smooth ellipse. The exception is that of the largest sweep angle tested at $\mathit{\Lambda}$ = 60, where undulations delayed the transition to a different wake structure. 

The results indicate that undulated cylinders with sweep angles of $\mathit{\Lambda} = 30$ or smaller can effectively reduce RMS lift and the resulting vortex-induced vibration reduction, and to a lesser extent, can reduce drag force. Observations made at two bioinspired Reynolds number values demonstrated that sweep-induced force suppression and wake modifications are not strongly dependent on Reynolds number within this range. While higher Reynolds number flows exhibit increased unsteadiness and less coherent wake structures, the primary effects of surface undulations, suppression of vortex-induced RMS lift and convergence towards ellipse-like behavior with increasing sweep angle, persisted across both $Re$ values. This demonstrates that the hydrodynamic benefits of the whisker-inspired geometry are robust to realistic changes in swimming speed. 
However, analysis of span-averaged turbulent kinetic energy reveals that the relative wake energy reduction provided by the whisker geometry becomes more pronounced at $Re = 500$, particularly at the largest sweep angles. This indicates that although force suppression diminishes with sweep, the whisker continues to alter wake structure in a manner that limits the persistence of energetic turbulent structures more effectively than an ellipse. Thus, wake stabilization effects of the undulated geometry extend beyond the range of effective force suppression which became comparable to an ellipse at a sweep angle of 45 degrees. This convergence reflects the balance between sweep-induced attenuation of forces and the diminishing influence of whisker-induced suppression, rather than a sharply defined biological threshold. 
In a real whisker array, which spans a continuum of effective orientations, this crossover would be distributed across whiskers and along their length. However, whiskers are more likely to occupy the range of orientations associated with stronger lift suppression when protracted forward into the flow, suggesting this behavior may enhance the effectiveness of their sensing abilities.

While the present study does not model whisker motion, upstream wakes, or fluid-structure interactions, the observed RMS lift trends suggest likely implications for VIV and hydrodynamic sensing. At small sweep angles, undulations strongly disrupt coherent vortex shedding, reducing unsteady forces and indicating the greatest benefit for vibration suppression. As sweep increases, undulation effects diminish and the wake becomes more coherent, but the absolute magnitude of unsteady forces is lower, so background forcing remains moderate. These results imply that whiskers achieve the greatest VIV reduction and potentially enhanced sensing when protracted forward. The persistence of suppression across a broader range of orientations highlights functional robustness to natural variation in whisker angle and environmental flow. By reducing background unsteady forces, whiskers remain relatively still in mean flow, which would allow incoming disturbances from upstream wakes or objects to produce more noticeable perturbations, consistent with theories that this mechanism underlies their effectiveness as hydrodynamic sensors and that this effect persists with the addition of sweep angle.

These findings highlight the importance of sweep angle as a biologically-informed and physically significant parameter. Quantitative assessment of VIV and sensing performance will require further investigation incorporating flexible whiskers or interactions with upstream objects, and the extension of this geometry and its beneficial qualities to higher Reynolds numbers remains an open question.

\section{Acknowledgments}
The authors acknowledge funding support by the National Science Foundation Award CBET-2035789 (JAF) and CBET-2037582 (RBC), the Naval Research Enterprise Internship Program, and the Department of Defense SMART Scholarship Program. This research was conducted with computational resources at the Department of Defense High Performance Computing Modernization Program and the University of Wisconsin-Madison Center for High Throughput Computing.

We thank Dr. Colleen Reichmuth, the Pinniped Cognition and Sensory Systems Laboratory and Thia Elliott for imagery of the harbor seal used in Figure~\ref{fig:Seal_photo}, and Sprouts the seal for the insight and inspiration that he provided to this research.

DOD Distribution Statement A. Approved for public release: distribution is unlimited.

\appendix
\section{Mesh Independence}\label{appA}
The numerical grid used for these simulations consists of 4.14 million cells and is the same as the mesh developed and validated in previous literature \cite{lyons_flow_2020, yuasa_simulations_2021, lyons_wavelength_2022} for simulations at $Re$ = 500 after comparison to structured meshes containing 1, 2, and 6 million cells. Figure~\ref{fig:meshvalid} shows that, while still well-resolved, the addition of sweep to the domain produces a larger spanwise component of flow for which the original mesh was not validated. 

To ensure that mesh resolution is sufficient to capture physical flow features and align results with those expected from prior literature, a series of mesh resolution levels were simulated at the largest sweep angle of $\mathit{\Lambda}$ = 60 (with its larger spanwise velocity component) to compare their force values and flow field characteristics. The dimensions of the meshes used and resulting values for mean drag coefficient and shedding frequency are given in Table~\ref{tab:meshtable}. Also shown below in Figure~\ref{fig:meshvalid} are contours of time-averaged Reynolds stress $\overline{u'u'}$ at equivalent spanwise locations for each mesh. Given the mesh candidates tested, the variation in forces and flow field features is minimal at and above the 4.14M cell mesh for the worst-case scenario spanwise flow condition. This mesh resolution is therefore chosen once again to minimize computational cost while delivering accurate results for the metrics of interest in this investigation.

\setcounter{figure}{0}
\renewcommand{\thefigure}{A\arabic{figure}}
\begin{figure}[hbt!]
\centering
\begin{subfigure}[b]{0.32\textwidth}
\centering
\includegraphics[width=\textwidth]{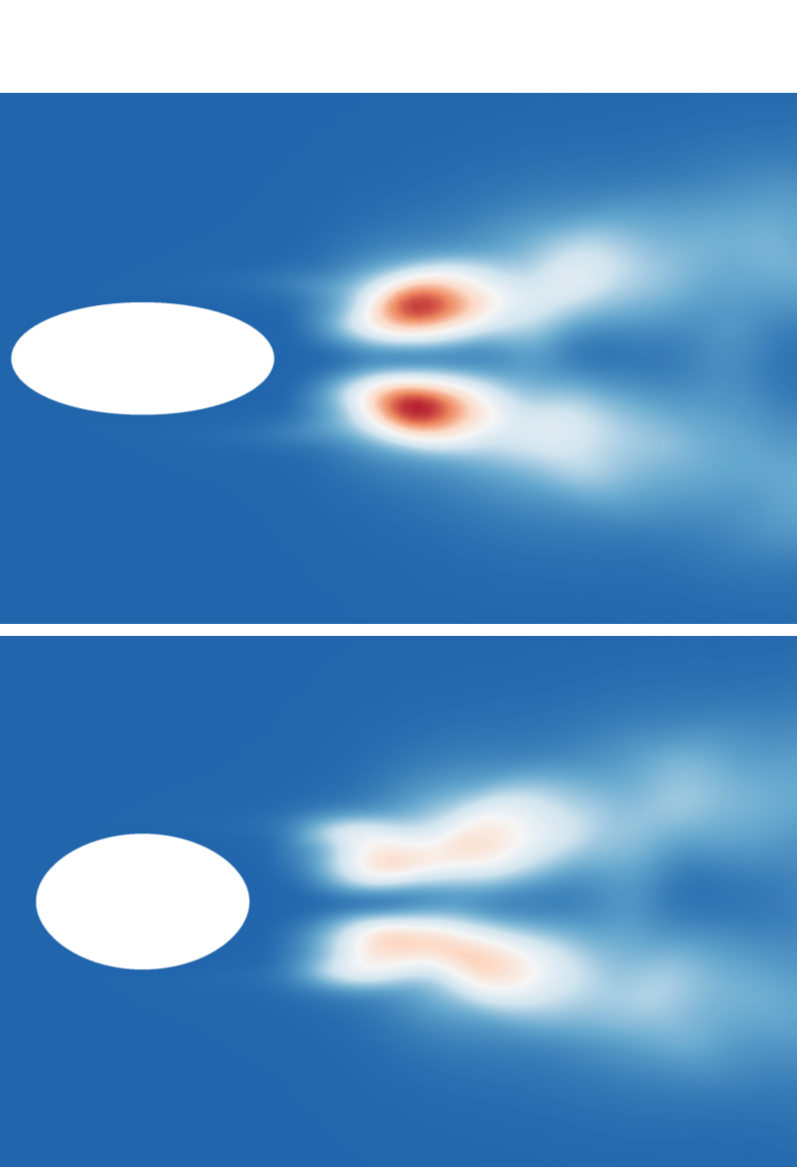}
\caption{$\bm{N_{total}}$ = 4.14M}
\label{fig:a1a}
\end{subfigure}
\centering
\begin{subfigure}[b]{0.32\textwidth}
\centering
\includegraphics[width=\textwidth]{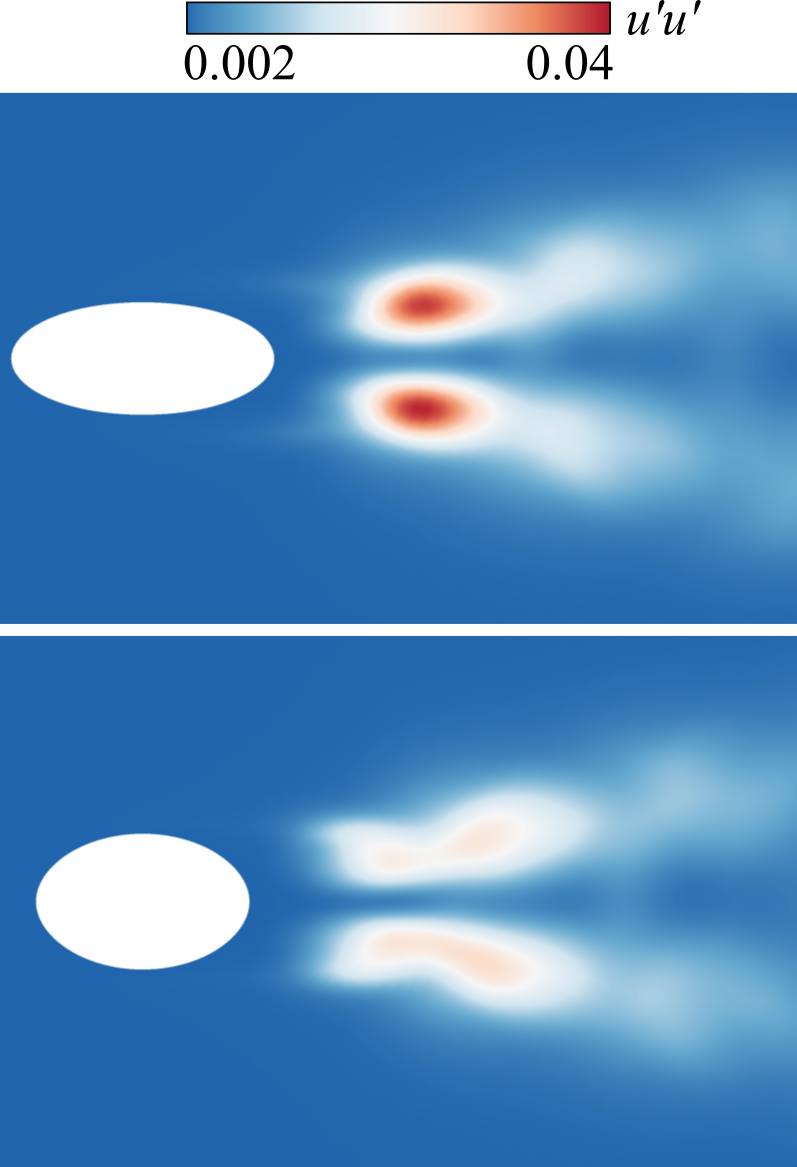}
\caption{$\bm{N_{total}}$ = 6.21M}
\label{fig:a1b}
\end{subfigure}
\centering
\begin{subfigure}[b]{0.32\textwidth}
\centering
\includegraphics[width=\textwidth]{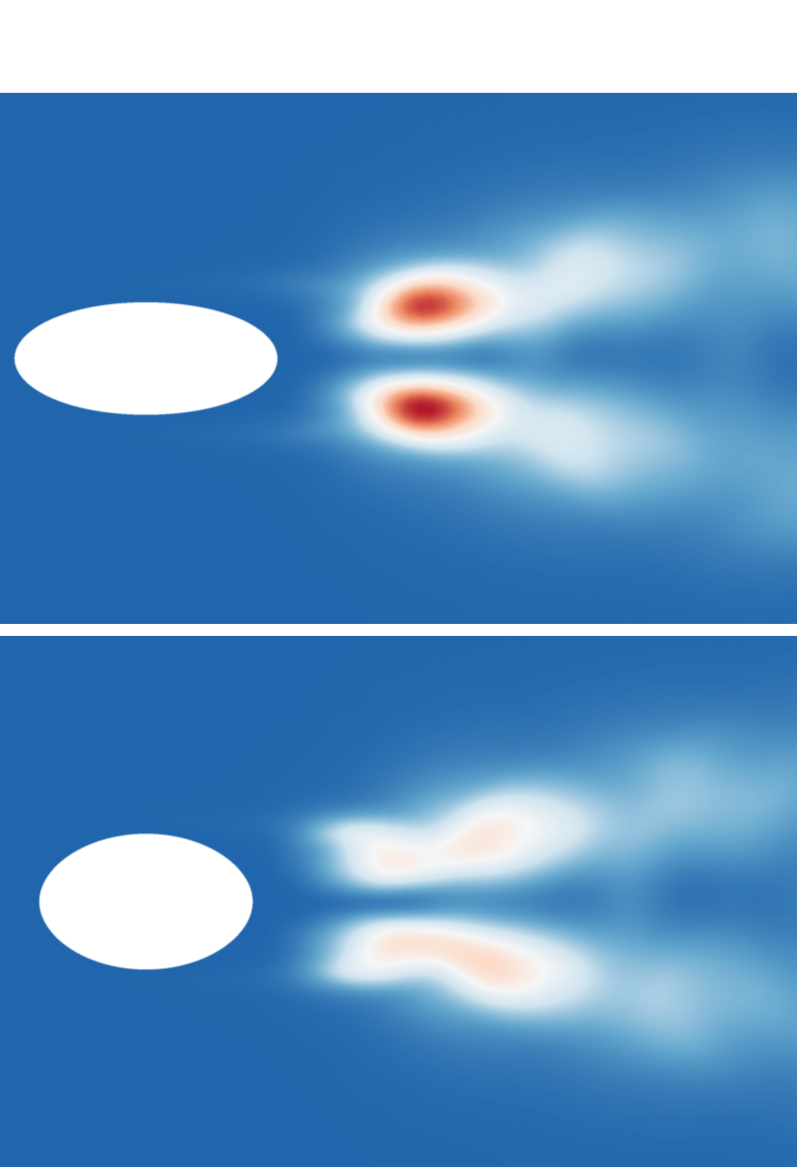}
\caption{$\bm{N_{total}}$ =  8.28M}
\label{fig:a1c}
\end{subfigure}
\caption{Comparison of time-averaged Reynolds stress $\bm{\overline{u'u'}}$ on equal colorbars for candidate meshes at locations of $\bm{z/\lambda}$ = 1 (top) and $\bm{z/\lambda}$ = 0.5 (bottom) along the span.}
\label{fig:meshvalid}
\end{figure}

\setcounter{table}{0}
\renewcommand{\thetable}{A\arabic{table}}
\begin{table}[hbt!] 
\rowcolors{2}{gray!25}{white}
\rowcolors{2}{gray!25}{white}
\rowcolors{2}{gray!25}{white}
\caption{Mesh parameters for each tested grid with resulting force and shedding frequency values. The dimensions listed are total nodes $\bm{N_{total}}$, nodes in the spanwise direction $\bm{N_z}$, azimuthal direction $\bm{N_\theta}$, radial direction $\bm{N_r}$, and the resulting cell sizes in the spanwise direction $\bm{\Delta z/T}$, and minimum in the radial direction $\bm{\Delta r/T_{min}}$.}
\label{tab:meshtable}
\centering
\begin{tabular}{|c|c|c|c|c|c|c|c|}
\hline
$N_{total}$ & $N_z$   & $N_\theta$ & $N_r$ & $\Delta z/T$ & $\Delta r/T_{min}$ & $\overline{C_D}$ & $St$\\ \hline \hline
4.14M & 170 & 160 & 154 & 0.041 & 0.003 & 0.236 & 0.1067\\ \hline
6.21M & 255 & 160 & 154 & 0.027 & 0.003 & 0.234 & 0.1067\\ \hline
8.28M & 340 & 160 & 154 & 0.020 & 0.003 & 0.234 & 0.1067\\ \hline 
\end{tabular}
\end{table}

\clearpage
\bibliography{reference.bib}

\end{document}